\documentclass[journal,12pt,letterpaper,draftcls,onecolumn]{IEEEtran}%
\usepackage{amsmath}
\usepackage{amsfonts}
\usepackage{cite}
\usepackage{amssymb}
\usepackage{mathrsfs}
\usepackage{graphicx}
\usepackage[usenames,dvipsnames]{color}
\usepackage{epstopdf}
\usepackage[all]{xy}
\usepackage[]{mcode}
\usepackage{supertabular}
\usepackage{subcaption}
\usepackage{array}
\usepackage{algorithmic}
\usepackage{algorithm}

\providecommand{\U}[1]{\protect\rule{.1in}{.1in}}
\pdfoutput=1
\providecommand{\U}[1]{\protect\rule{.1in}{.1in}}

\newenvironment{example}[1][Example]{\begin{trivlist}
\item[\hskip \labelsep {\bfseries #1}]}{\end{trivlist}}

\newcommand{\qed}{\nobreak \ifvmode \relax \else
      \ifdim\lastskip<1.5em \hskip-\lastskip
      \hskip1.5em plus0em minus0.5em \fi \nobreak
      \vrule height0.75em width0.5em depth0.25em\fi}

\begin{document}

\title{Enhanced Huffman Coded OFDM with \\Index Modulation}
\author{Shuping Dang, \textit{Member, IEEE}, Shuaishuai Guo, \textit{Member, IEEE}, \\Justin P. Coon, \textit{Senior Member, IEEE},  Basem Shihada, \textit{Senior Member, IEEE}, \\Mohamed-Slim Alouini, \textit{Fellow, IEEE}
  \thanks{This work was supported by KAUST Office of Sponsored Research.
  
  S. Dang, B. Shihada, and M.-S. Alouini are with Computer, Electrical and Mathematical Sciences and Engineering Division, King Abdullah University of Science and Technology (KAUST), 
Thuwal 23955-6900, Saudi Arabia (e-mail: \{shuping.dang, basem.shihada, slim.alouini\}@kaust.edu.sa).

S. Guo was with the Computer, Electrical and Mathematical Sciences and Engineering Division, King Abdullah University of Science and Technology (KAUST), Thuwal 23955-6900, Saudi Arabia. He is now with Shandong Provincial Key Laboratory of Wireless Communication Technologies and School of Control Science and Engineering, Shandong University, Jinan 250061, P.R. China (e-mail: shuaiguosdu@gmail.com).

J.P. Coon is with the Department of Engineering Science, University of Oxford, Oxford OX1 3PJ, U.K. (e-mail: justin.coon@eng.ox.ac.uk)
}}

\maketitle

\begin{abstract}
In this paper, we propose an enhanced Huffman coded orthogonal frequency-division multiplexing with index modulation (EHC-OFDM-IM) scheme. The proposed scheme is capable of utilizing all legitimate subcarrier activation patterns (SAPs) and adapting the bijective mapping relation between SAPs and leaves on a given Huffman tree according to channel state information (CSI). As a result, a dynamic codebook update mechanism is obtained, which can provide more reliable transmissions. We take the average block error rate (BLER) as the performance evaluation metric and approximate it in closed form when the transmit power allocated to each subcarrier is independent of channel states. Also, we propose two CSI-based power allocation schemes with different requirements for computational complexity to further improve the error performance. Subsequently, we carry out numerical simulations to corroborate the error performance analysis and the proposed dynamic power allocation schemes. By studying the numerical results, we find that the depth of the Huffman tree has a significant impact on the error performance when the SAP-to-leaf mapping relation is optimized based on CSI. Meanwhile, through numerical results, we also discuss the trade-off between error performance and data transmission rate and investigate the impacts of imperfect CSI on the error performance of EHC-OFDM-IM. 
\end{abstract}


\section{Introduction}
\IEEEPARstart{T}{o} cope with the emerging requirements of high reliability and high transmission rate in next generation networks, novel modulation schemes have been regarded as promising solutions and have been exhaustively studied in recent years \cite{8006223,8070115,8315127,8417419}. One of the most noticeable directions is index modulation (IM), which is related to permutation modulation and parallel combinatory  modulation but developed in an independent manner since 2001 \cite{8315127}. In recent decades, with the proliferation of multi-antenna and multi-carrier systems, IM has found its suitable application scenarios, and has attracted researchers' attention with proposals of spatial modulation (SM) and orthogonal frequency-division multiplexing with IM (OFDM-IM), which are applications of IM in the spatial and frequency domains \cite{8004416,8750780}. Because SM is mainly designed for multi-antenna systems, which demands relatively high system complexity and a large device size \cite{8353362}, OFDM-IM has become more popular by releasing these constraints since its proposal in 2013 \cite{6587554}. Another advantage of OFDM-IM is that the mature standards, protocols, and existing communication infrastructure for OFDM in fourth generation (4G) wireless networks can be applied to OFDM-IM systems without a considerable amendment, because of the similarity between both paradigms in practice \cite{8101465}. Rigorous studies have proved the performance superiority of OFDM-IM over plain OFDM without IM in terms of error performance and transmission efficiency \cite{7330022,7469311}. The solid theoretical basis and the easy-to-implement attribute pave the way for OFDM-IM to prosper in next generation networks, so as to meet the stringent requirements of high reliability and high transmission rate.

Apart from the aforementioned benefits of OFDM-IM, there is one question that has accompanied OFDM-IM since its proposal until now: what is an efficient way to design a codebook for OFDM-IM that maps the codewords to subcarrier activation patterns (SAPs) in a bijective manner? Even before the proposal of OFDM-IM, two cognate schemes called subcarrier-index modulation (SIM) OFDM\footnote{Because of the shared nature, SIM OFDM is also regarded as an interchangeable term with OFDM-IM in some papers.} and enhanced SIM OFDM have paid special attention to this crucial question \cite{5449882,6162549}. In \cite{5449882}, a forward error control mechanism was introduced to realize a heuristic mapping procedure, which raises the system complexity unnecessarily, and in \cite{6162549}, the coding efficiency was significantly reduced, albeit circumventing the forward error control mechanism. The classic OFDM-IM proposed in \cite{6587554} adopts a truncated subset of SAPs with a fixed number of active subcarriers to match fixed-length bit sequences, which still decreases a portion of the transmission rate. In \cite{8241721}, a combined mapping relation with variable numbers of active subcarriers was proposed, which increases the transmission rate at the cost of increased system complexity. To reduce the system complexity raised by the combined OFDM-IM, another mapping relation relying on a special always-on subcarrier was studied in \cite{8361430,8358694,8476574,8614439}, by which the number of active subcarriers also varies. Such a feature of variable numbers of active subcarriers would cause practical issues in currently existing infrastructure and is thereby undesirable.

To mitigate the dilemma between performance and complexity, Huffman coding is useful to construct a bijective mapping relation between SAPs consisting of a fixed number of active subcarriers to variable-length bit sequences \cite{4051119}. Huffman coding was introduced to two spatial applications of IM in \cite{6253219} and \cite{7933242}, which studied the effects of the architecture of the Huffman tree on the system performance. According to the feedback from the receiving side, the transmitter is capable of dynamically updating the codebook to maximize the transmission rate. When it comes to OFDM-IM, a mapping relation between variable-length bit sequences and SAPs was defined in \cite{7370760}, by which all legitimate SAPs can be employed in the codebook, and a low-complexity detection scheme for OFDM-IM systems is thereby enabled. Although Huffman coding has not been explicitly mentioned in this work, the proposed system applies the core principle of Huffman coding and selects the Huffman tree with the minimum depth to encode incoming bits. The optimization of Huffman coded OFDM-IM by adjusting the architecture of the Huffman tree was formulated in \cite{8279383} and \cite{8588988} with various optimization objectives. They both provide good thoughts on how the Huffman tree architecture could be adapted to provide a higher capacity and transmission rate when utilizing all legitimate SAPs. In \cite{8469479}, Huffman coding was applied to OFDM-IM with the help of channel state information (CSI). However, the Huffman coded OFDM-IM scheme proposed in \cite{8469479} did not fully exploit all legitimate SAPs, and thereby only shows limited improvements. The latest research results with respect to the optimization of the achievable rate of binary-tree coded OFDM-IM (a generic version of Huffman coded OFDM-IM) were presented in \cite{8704951}. In this work, the achievable rate is optimized by dynamically varying the SAP probability distribution and transmit power allocated to active subcarriers.

To attain a proper transmission rate by involving all legitimate SAPs in a codebook and dynamically update the codebook by CSI to enhance the error performance, we propose an enhanced Huffman coded OFDM-IM (EHC-OFDM-IM) scheme in this paper. In order to clarify the contributions of this paper and the novelty compared to prior works, we detail the difference between the proposed EHC-OFDM-IM scheme with other existing and related schemes as follows:
\begin{itemize}
\item Different from the classic Huffman coded OFDM-IM scheme proposed in \cite{7370760} that does not rely on any channel feedback, we employ CSI at the transmitting side to provide an adaptive optimization mechanism.
\item Different from the adaptive Huffman coded OFDM-IM schemes proposed in \cite{8279383} and \cite{8588988} that focus on the optimization of the architecture of the Huffman tree, we instead focus on the optimization of the bijective mapping relation between all legitimate SAPs and leaves on a \textit{given} Huffman tree, i.e., the variable-length bit sequences. Therefore, EHC-OFDM-IM can be applied on top of the optimized systems constructed in \cite{8279383} and \cite{8588988} to further enhance their error performance. 
\item Different from the adaptive Huffman coded OFDM-IM scheme proposed in \cite{8469479} that does not exploit all SAPs but only involves a binary probability vector to pair up a portion of SAPs based on channel qualities, EHC-OFDM-IM includes all legitimate SAPs in the codebook and use them for index bit transmission.
\item Different from the binary-tree coded OFDM-IM proposed in \cite{8704951} that undertakes the task of optimizing the achievable rate subject to the availability of channel knowledge at the transmitter, we focus on the task of improving the error performance, which formulates a different optimization problem and employs a different optimization strategy.
\item We borrow the SAP ordering principle given in \cite{8519769} to our paper to construct the pairing relation between SAPs to the leaves on a given Huffman tree. Although both schemes employ the CSI at the transmitter to perform codebook optimization, the system model and optimization scenario in this paper are different from the OFDM-IM aided by lexicographic codebook design (LCD-OFDM-IM) proposed in \cite{8519769}. In particular, the LCD-OFDM-IM scheme selects a set of SAPs by the priority and excludes the other SAPs in the codebook, which does not fully exploit the largest span of legitimate SAPs but could be able to harvest a diversity gain. The EHC-OFDM-IM scheme employs the priority of each SAP to pair it up to a leaf on the Huffman tree. In this way, the SAP with higher priority will be used more frequently, so as to provide better error performance with a coding gain. By sacrificing the diversity gain, the EHC-OFDM-IM scheme utilizes all legitimate SAPs and thereby has a higher rate.
\end{itemize}
In order to provide a comprehensive and qualitative comparison so as to emphasize the contributions of the proposed EHC-OFDM-IM in an explicit manner\footnote{Recently, an open-source IM toolkit has been released aiming at facilitating the quantitative comparisons among multiple IM schemes and accelerating open innovation in IM studies. Interested readers can refer to \cite{8759857} for more details.}, we list the key attributes of various OFDM-IM and sibling schemes in Table \ref{compare}.

\begin{table*}[!t]
\renewcommand{\arraystretch}{1.3}
\caption{Comparisons and key attributes of various OFDM-IM and sibling schemes.}
\label{compare}
\centering
\scalebox{0.8}{
\begin{tabular}{c|c|c|c|c|c|c|c|c}
\hline
Scheme & Literature & \begin{tabular}{@{}c@{}} Variable no.\\  of active \\ subcarriers? \end{tabular} & \begin{tabular}{@{}c@{}} Variable bit \\stream length? \end{tabular} & \begin{tabular}{@{}c@{}} CSI required\\at the Tx? \end{tabular} & \begin{tabular}{@{}c@{}} Adaptive\\codebook? \end{tabular} &  \begin{tabular}{@{}c@{}} All legit. \\SAPs used? \end{tabular} & \begin{tabular}{@{}c@{}} Diversity gain\\ harvested? \end{tabular} & \begin{tabular}{@{}c@{}} Coding gain\\ harvested? \end{tabular}\\
\hline
\hline
\begin{tabular}{@{}c@{}} EHC-OFDM-IM\\ (proposed) \end{tabular}  & -- & NO & YES & YES & YES & YES & NO & YES \\
\hline
\begin{tabular}{@{}c@{}} Classic Huffman\\ coded OFDM-IM \end{tabular} & \cite{7370760} & NO & NO & NO & NO & YES & NO & NO \\
\hline
\begin{tabular}{@{}c@{}} Adaptive Huffman\\coded OFDM-IM I \end{tabular} & \cite{8279383} & NO & YES & YES & YES & YES & NO & YES \\
\hline
\begin{tabular}{@{}c@{}} Adaptive Huffman\\coded OFDM-IM II \end{tabular} & \cite{8588988} & NO & YES & NO & YES & YES & NO & NO \\
\hline
\begin{tabular}{@{}c@{}} Adaptive Huffman\\coded OFDM-IM III \end{tabular} & \cite{8469479} & NO & YES & YES & YES & NO & NO & YES \\
\hline
\begin{tabular}{@{}c@{}} Binary-tree coded\\ OFDM-IM \end{tabular} & \cite{8704951} & NO & YES & YES & YES & YES & NO & YES \\
\hline
\begin{tabular}{@{}c@{}} Classic OFDM-IM \end{tabular} & \cite{6587554} & NO & NO & NO & NO & NO & NO & NO \\
\hline
\begin{tabular}{@{}c@{}} Adaptive\\OFDM-IM \end{tabular} & \cite{8241721,8269167} & YES & YES & YES & YES & NO & YES & YES \\
\hline
\begin{tabular}{@{}c@{}} OFDM-IM with\\an always-on\\subcarrier\end{tabular} & \cite{8361430,8358694,8476574,8614439} & YES & YES & NO & NO & NO & NO & NO \\
\hline
\begin{tabular}{@{}c@{}} LCD-OFDM-IM\end{tabular} & \cite{8519769} & NO & NO & YES & YES & NO & YES & YES \\
\hline
\begin{tabular}{@{}c@{}} SIM OFDM \end{tabular} & \cite{5449882} & YES & YES & NO & YES & NO & NO & NO \\
\hline
\begin{tabular}{@{}c@{}} Enhanced\\SIM OFDM \end{tabular} & \cite{6162549} & NO & NO & NO & NO & YES & NO & NO \\
\hline
\begin{tabular}{@{}c@{}} OFDM-SNM\end{tabular} & \cite{8362748} & YES & YES & NO & NO & NO & NO & NO \\
\hline
\begin{tabular}{@{}c@{}} Enhanced\\OFDM-SNM\end{tabular} & \cite{8703169} & YES & YES & YES & YES & NO & NO & YES \\
\hline
\begin{tabular}{@{}c@{}} FQAM/FPSK \end{tabular} & \cite{6824956,7763523} & NO & NO & NO & NO & NO & NO & NO \\
\hline
\end{tabular}}
\end{table*}

Following the proposal of EHC-OFDM-IM, we analyze its error performance when maximum-likelihood (ML) detection is in use and the transmit power allocated to each active subcarrier does not depend on CSI, and we approximate the average block error rate (BLER) in closed form by the union bound and an exponential approximation of the Q-function.

Apart from the optimization of the mapping relation between SAPs and variable-length bit sequences, power allocation is another key technique that could be used to enhance error performance. Although there have existed a variety of power allocation schemes proposed for plain OFDM \cite{1097882,1542594,4232711,5208735,6148193}, for OFDM-IM, there are only a limited number of works dedicated to power allocation. The power allocation for OFDM-IM systems was first investigated in \cite{8108623}, in which two power allocation schemes aiming at optimizing error performance were proposed based on a simplistic approximation counting only two kinds of pairwise error events. Profile-based power allocation for OFDM-IM was studied in \cite{8108528}, which resorts to the maximization of the minimum Euclidean distance (MED). Better error performance is attainable with reasonable computational complexity. Power allocation problems for adaptive cooperative OFDM-IM and dual-mode OFDM-IM were formulated and discussed in \cite{8269167} and \cite{8051264}, respectively. The power allocation problem for maximizing the rate of binary-tree coded OFDM-IM  was also investigated in \cite{8704951}. To further enhance the error performance of EHC-OFDM-IM, we study the corresponding power allocation problems and propose two suboptimal power allocation schemes based on the CSI at the transmitter with different requirements for computational complexity, which are capable of achieving a lower average BLER and are also easy to implement.

Numerical results generated by Monte Carlo simulations are provided to verify the error performance superiority and the error performance analysis of EHC-OFDM-IM either with or without applying the CSI-based power allocation schemes. Furthermore, we also investigate the trade-off in the architecture of the Huffman tree between error performance and data transmission rate as well as the impacts of imperfect CSI on the error performance through the numerical results provided.  To study the bitwise error event in the context of variable-length bit sequences by EHC-OFDM-IM, we design an appropriate way to count the number of bit errors and numerically investigate the bit error rate (BER) by numerical results.  The proposed scheme and algorithms in this paper can be easily extended to more sophisticated configurations, e.g., space-time-frequency IM, dual-mode OFDM-IM, and multi-mode OFDM-IM \cite{8322306,7547943,7936676}.

The rest of the paper is organized as follows. In Section \ref{sm}, we present the system model. We then give the details on how Huffman coding can be enhanced with CSI at the transmitter in Section \ref{ehc}. Then, we carry out error performance analysis in Section \ref{epa} and investigate the power allocation problems in Section \ref{pas}. Following the analysis, numerical results are illustrated, compared, and discussed in Section \ref{nr}. Finally, the paper is concluded in Section \ref{c}.

\section{System Model}\label{sm}
Although a full set of subcarriers will normally be separated into different groups by OFDM-IM, for simplicity, we now only focus on a single group of $N$ subcarriers in this paper. For each transmission attempt, $K$ out of these $N$ subcarriers are chosen to be activated so as to represent an SAP for conveying extra index bits, which are assumed to be equiprobable. Therefore, the number of all legitimate SAPs is $S=\binom{N}{K}$, where $\binom{\cdot}{\cdot}$ is the binomial coefficient. Further assume that $M$-ary phase-shift keying ($M$-PSK) is in use to modulate equiprobable data bits due to its \textit{constant modulus nature} and \textit{rotational symmetry} \cite{7330022,7054523}. We write the transmitted OFDM block as
\begin{equation}
\mathbf{x}(s,\mathbf{m})=[x(s,m_1),x(s,m_2),\dots,x(s,m_N)]^T\in\mathbb{C}^{N\times 1},
\end{equation}
where $s\in\mathcal{S}=\{1,2,\dots,S\}$ is the index of the SAP; $\mathbf{m}=[m_1,m_2,\dots,m_N]\in\mathcal{M}^{N}=\{0,1,2,\dots,M\}^N$ is the joint index vector of $K$ $M$-ary data symbols conveyed on active subcarriers and $N-K$ nulls on inactive subcarriers; $x(s,m_n)$ is a normalized data symbol only if the $n$th subcarrier is active, or null otherwise\footnote{For convenience, we let $m_n=0$ indicate that the $n$th subcarrier is inactive when the $s$th SAP is adopted for transmission.}; $(\cdot)^T$ denotes the transpose operation on the enclosed vector/matrix.

Assuming perfect synchronization, pulse shaping, equalization, sampling and knowledge of CSI, the received OFDM block after sampling and discarding the cyclic prefix can be expressed as \cite{1542594}
\begin{equation}
\begin{split}
\mathbf{y}(s,\mathbf{m})&=[y(s,m_1),y(s,m_2),\dots,y(s,m_N)]^T\\
&=\mathbf{\Omega}(s)\mathbf{H}\mathbf{x}(s,\mathbf{m})+\mathbf{w}\in\mathbb{C}^{N\times 1},
\end{split}
\end{equation}
where $\mathbf{\Omega}(s)=\mathrm{diag}\{\sqrt{P_t(s,1)},\sqrt{P_t(s,2)},\dots,\sqrt{P_t(s,N)}\}$ and $P_t(s,n)$ is the transmit power allocated to the $n$th subcarrier in the $s$th SAP\footnote{For simplicity, we tentatively suppose that $P_t(s,n)$ is independent of CSI and has been set when analyzing the error performance of EHC-OFDM-IM systems. To further enhance the error performance, CSI-based power allocation can be applied, which, however, can not be analyzed thoroughly to the best of authors' knowledge. Details of the CSI-based power allocation can be found in Section \ref{pas}.}; $\mathbf{H}=\mathrm{diag}\{h(1),h(2),\dots,h(N)\}$ and $\mathbf{w}=[w(1),w(2),\dots,w(N)]^{T}$ represent the diagonal channel state matrix and the additive white Gaussian noise (AWGN) vector, whose entries are the channel coefficients and AWGN samples complying with complex Gaussian distributions $\mathcal{CN}(0,\mu)$ and $\mathcal{CN}(0,N_0)$, respectively; $\mu$ is the average channel power gain, and $N_0$ is the average noise power. Note that, all channel coefficients are assumed to be independent, which can be easily achieved by interleaved subcarrier grouping \cite{6841601}. Here, we adopt the Rayleigh fading model, and all channel power gains $G(1),G(2),\dots,G(N)$ ($G(n)=|h(n)|^2$, $\forall~n\in\mathcal{N}=\{1,2,\dots,N\}$) are assumed to be independently and exponentially distributed with the same average channel power gain $\mu$, which have the probability density function (PDF) and cumulative distribution function (CDF) given by \cite{7577711}
\begin{equation}
f_G(\epsilon)=\frac{1}{\mu}\mathrm{exp}\left(-\frac{\epsilon}{\mu}\right)\Leftrightarrow F_G(\epsilon)=1-\mathrm{exp}\left(-\frac{\epsilon}{\mu}\right).
\end{equation}

Receiving the contaminated OFDM block, the receiver can utilize ML detection scheme to estimate the transmitted OFDM block by the criterion \cite{6841601}
\begin{equation}\label{sdakskjdk2dmldetec}
\begin{split}
&\hat{\mathbf{x}}(\hat{s},\hat{\mathbf{m}})=\underset{\dot{\mathbf{x}}(\dot{s},\dot{\mathbf{m}})\in\mathcal{X}}{\arg\min}\begin{Vmatrix}\mathbf{y}(s,\mathbf{m})-\mathbf{\Omega}(\dot{s})\mathbf{H}\dot{\mathbf{x}}(\dot{s},\dot{\mathbf{m}})\end{Vmatrix}_F,
\end{split}
\end{equation}
where $\begin{Vmatrix}\cdot\end{Vmatrix}_F$ denotes the Frobenius norm of the enclosed vector/matrix, and $\mathcal{X}$ is the set of all legitimate transmit OFDM blocks, which has a cardinality of $SM^K$. 

When the OFDM block $\mathbf{x}(s,\mathbf{m})$ is transmitted, the corresponding BLER can be written as\footnote{Although in most publications associated with OFDM-IM, bit error rate (BER) is adopted as an error performance evaluation metric, BER is intuitive and more suited for measuring the error performance when bit streams with a fixed and equal length are transmitted. However, by involving Huffman coding, the length of the transmitted bit stream varies when different SAPs are chosen. When variable-length bit streams are taken into consideration, we cannot `intuitively' determine the number of bit errors without stipulating a unified quantifying rule. However, defining such a rule is not a trivial task. The quantifying rule must be general enough to cover all possible scenarios, and in the meantime aligned with common sense, e.g., the number of bit errors cannot be higher than the number of transmitted bits. Moreover, the rule must also lead to consistent evaluations for all scenarios. Therefore, to avoid unnecessary discussion regarding how to count bit errors when transmitted and erroneously decoded bit streams have different lengths, we adopt BLER as the error performance evaluation metric in this paper.} 
\begin{equation}\label{blerasjdh2}
P_e(\mathbf{x}(s,\mathbf{m}))=\mathbb{P}\left\lbrace\hat{\mathbf{x}}(\hat{s},\hat{\mathbf{m}})\neq \mathbf{x}(s,\mathbf{m})\right\rbrace,
\end{equation}
where $\mathbb{P}\left\lbrace\cdot\right\rbrace$ denotes the probability of the random event enclosed. Consequently, by considering all legitimate transmit OFDM blocks, we can write the average BLER as
\begin{equation}\label{pingjundasebls}
\bar{P}_e=\underset{\mathbf{x}(s,\mathbf{m})\in\mathcal{X}}{\mathbb{E}}\left\lbrace P_e(\mathbf{x}(s,\mathbf{m}))\right\rbrace,
\end{equation}
where $\mathbb{E}\left\lbrace\cdot\right\rbrace$ denotes the expected value of the enclosed random variable.

\section{Enhanced Huffman Coding with CSI}\label{ehc}
By introducing Huffman coding, a full binary tree with $S$ leaves representing $S$ SAPs is constructed, which is also termed the Huffman tree. Here, we are only interested in the full binary tree because of its \textit{prefix-free property} and \textit{instantaneous decodability} \cite{yang2003h}. For a full binary tree with $S$ leaves, its architecture can be uniquely characterized by a set of depths $\{d(s)\}_{s=1}^S$ or equivalently a set of occurrence probabilities $\{p(s)\}_{s=1}^S$, where $p(s)=\left(\frac{1}{2}\right)^{d(s)}$ and $\sum_{s=1}^{S}p(s)=1$ \cite{8279383}. The number of full binary trees with $S$ leaves is the Catalan number $C(S)=\frac{1}{(S+1)}\binom{2S}{S}$, among which the minimum and maximum depths are $\lfloor\log_2(S-1)\rfloor+1$ and $S-1$, respectively \cite{8704951}. Once the architecture of the Huffman tree used for Huffman encoding is given, the average encoding rate is fixed, which can be calculated by \cite{8588988}
\begin{equation}\label{sdjashdj2bbbbb}
B=\underbrace{\sum_{s=1}^{S}\left[\left(\frac{1}{2}\right)^{d(s)}d(s)\right]}_{B_I}+\underbrace{K\log_2(M)}_{B_D},
\end{equation}
where $B_I$ is the encoding rate of the index bits modulated by the SAP in units of bit per channel use (bpcu), and $B_D$ is the encoding rate of the data bits modulated by the data constellation symbols conveyed on $K$ active subcarriers in units of bpcu. 

Meanwhile, with CSI at the transmitter, there is still a possibility to further enhance the error performance by designing a bijective mapping relation between $S$ SAPs and $S$ leaves on the generated Huffman tree. This is because the occurrence probabilities of SAPs represented by different leaves are different, as long as the Huffman tree is not \textit{perfect}\footnote{A perfect Huffman tree is defined as the tree, on which all interior nodes have two children and all resultant leaves are with the same depth \cite{2008data}.}.

To enable the error performance optimization by CSI at the transmitter with channel power gains $G(1),G(2),\dots,G(N)$, we order the subchannels by their channel power gains and write the ranked channel power gains in ascending order as\footnote{To be general, the relation between the channel power gains and ranked channel power gains can be explicitly expressed in the following way. Assuming the full set of channel power gains is denoted as $\mathbf{G}=\{G(1),G(2),\dots,G(N)\}$, we have $G_{\left\langle 1\right\rangle}(\xi_1)=\underset{\mathbf{G}}{\min}\{G(n)\}$, $G_{\left\langle 2\right\rangle}(\xi_2)=\underset{\mathbf{G}\setminus \{G_{\left\langle 1\right\rangle}(\xi_1)\}}{\min}\{G(n)\}$, $G_{\left\langle 3\right\rangle}(\xi_3)=\underset{\mathbf{G}\setminus \{G_{\left\langle 1\right\rangle}(\xi_1),G_{\left\langle 2\right\rangle}(\xi_2)\}}{\min}\{G(n)\}$, $\dots$, $G_{\left\langle N\right\rangle}(\xi_N)=\underset{\mathbf{G}\setminus \{G_{\left\langle 1\right\rangle}(\xi_1),G_{\left\langle 2\right\rangle}(\xi_2),\dots,G_{\left\langle {N-1}\right\rangle}(\xi_{N-1})\}}{\min}\{G(n)\}=\underset{\mathbf{G}}{\max}\{G(n)\}$.}
\begin{equation}
G_{\left\langle 1\right\rangle}(\xi_1)<G_{\left\langle 2\right\rangle}(\xi_2)<\dots<G_{\left\langle N\right\rangle}(\xi_N),
\end{equation}
which characterizes the utilization preference of all $N$ subcarriers, where $\xi_n$ is the index of the subcarrier with the $n$th smallest channel power gain. 

However, the performance of OFDM-IM systems is associated with an SAP with $K$ active subcarriers instead of a single active subcarrier; we have to derive the utilization preference of $S$ SAPs. To obtain the utilization preference of $S$ SAPs, we resort to the pattern ordering principle by reverse lexicographic order presented in \cite{8519769}. Then, we can rank the $S$ SAPs by priority according to the following steps:
\subsubsection{Generate the first and the last ASVs}
The first activation state vector (ASV) corresponding to the best SAP can be constructed by creating a $N\times 1$ zero vector and then replacing the rightmost $K$ zeros by ones. Mathematically, we have the initial ASV in the following form
\begin{equation}
\mathbf{v}_1=[\underbrace{0,0,\dots,0}_{N-K},\underbrace{1,1,\dots,1}_{K}]^T.
\end{equation}
Similarly, the last ASV corresponding to the worst SAP can be constructed by creating a $N\times 1$ zero vector and then replacing the leftmost $K$ zeros by ones. Mathematically, we have the initial ASV in the following form
\begin{equation}
\mathbf{v}_S=[\underbrace{1,1,\dots,1}_{K},\underbrace{0,0,\dots,0}_{N-K}]^T.
\end{equation}
Both are generated as the reference points for the other ASVs.

\subsubsection{Enumerate $S$ ASVs}
Apart from both ends, we also need to have the ASVs in the middle. When the number of subcarriers $N$ is small, the easiest way to enumerate all $S$ ASVs is to generate an expanded set of $2^N$ ASVs by releasing the requirement on the number of active subcarriers. We can thereby have the expanded set as
\begin{equation}
\begin{split}
\{&[0,0,\dots,0,0]^T, [0,0,\dots,0,1]^T\\
  &[0,0,\dots,1,0]^T, [0,0,\dots,1,1]^T\\
  &~~~~~~~~~~~~~~~~\vdots\\
  &[1,1,\dots,1,0]^T, [1,1,\dots,1,1]^T\}\\
\end{split}.
\end{equation}
Subsequently, we check the Hamming weight of each element in the expanded set and obtain a subset by removing the elements whose Hamming weights do not equal $K$. The obtained subset contains all $S$ ASVs corresponding to the legitimate SAPs with $K$ active subcarriers. Albeit simple, the above approach might not be applicable when $N$ is large, because it is demanding to have the register space for $2^N$ vectors. 

For large $N$, we propose a \textit{state shifting} approach based on the first ASV $\mathbf{v}_1$ to yield the set of $S$ ASVs. Imagining that `1's in an ASV are entities and there is nothing in the positions of `0', we shift the leftmost `1' to the left by one place and obtain the next ASV as $[0,0,\dots,1,0,1,\dots,1]^T$. Similarly, we shift the second leftmost `1' to the left by one place and obtain the next ASV as $[0,0,\dots,1,1,0,\dots,1]^T$. We subsequently repeat this process $K$ times until a \textit{consecutive} string of $K$ `1's emerging again, and have the $(K+1)$th ASV $[\underbrace{0,0,\dots,0}_{N-K-1},\underbrace{1,1,\dots,1}_{K},0]^T$. 
After this, we shift the leftmost `1' in the consecutive string to the left by one place and reset all other $K-1$ `1's starting back to the rightmost position in the ASV, which results in the $(K+2)$th vector $[\underbrace{0,0,\dots,0}_{N-K-2},1,0,0,\underbrace{1,1,\dots,1}_{K-1}]^T$. We repeat this process until the next emergence of a consecutive string of $K$ `1's and so on. Finally, we terminate the state shifting process when the last ASV $\mathbf{v}_S$ is obtained. In this way, we can obtain the set of $S$ ASVs corresponding to all legitimate SAPs with $K$ active subcarriers. 

\subsubsection{Rank $S$ ASVs by reverse lexicographic order}
By the ASV generating approaches introduced above, we can easily enumerate all legitimate SAPs by their ASVs $\mathbf{v}_s=[v_s(\xi_1),v_s(\xi_2),\dots,v_s(\xi_N)]^T\in\{0,1\}^{N\times 1}$, where $v_s(\xi_n)$ is either `0' or `1' depending on whether the $\xi_n$th subcarrier (i.e., the subcarrier with the $n$th smallest channel power gain) is activated\footnote{For example, $\mathbf{v}_s=[0,1,0,1]^T$ does not mean that the second and fourth subcarriers shall be activated by their indices, but implies that the subcarriers with the second and fourth smallest channel power gains, i.e., the second and fourth \textit{ordered} subcarriers, shall be activated.}. In particular, we rank all $S$ SAPs in ascending order by their ASVs:
\begin{equation}\label{2132131231555s}
\mathbf{v}_1<_{\mathsf{L}}\mathbf{v}_2<_{\mathsf{L}}\dots<_{\mathsf{L}}\mathbf{v}_S,
\end{equation}
where $<_{\mathsf{L}}$ is the lexicographic less-than sign, and we define $\mathbf{v}_s<_{\mathsf{L}}\mathbf{v}_t$ if $\omega_s<\omega_t$, where $\omega_s=\mathrm{b2d}(\mathbf{v}_s)$ and $\mathrm{b2d}(\cdot)$ converts a binary vector to a decimal number given the most significant digit on the left side. It has been explained in \cite{8519769} that the smallest SAPs in the lexicographic sense are more preferable in most cases when considering the error and outage performance for OFDM-IM systems with independent and identically distributed (i.i.d.) channel power gains. As a result, we rank the $S$ SAPs by priority according to the three steps given above.

In contrast to the classic OFDM-IM systems without Huffman coding \cite{8519769}, which simply select a subset of $2^{\lfloor\mathrm{log}_2(S)\rfloor}$ SAPs from the first to the $2^{\lfloor\mathrm{log}_2(S)\rfloor}$th  SAPs ranked by (\ref{2132131231555s}) and discarding the rest, Huffman coded OFDM-IM systems need to make full use of all $S$ SAPs in order to improve the data transmission rate. Although it would be impossible to harvest a diversity gain in Huffman coded OFDM-IM systems in this way, we can still exploit CSI at the transmitter and the ranked ASVs to obtain a coding gain. This is simply because, in classic OFDM-IM systems, all selected SAPs have the same probability of being activated, while in Huffman coded OFDM-IM systems, the occurrence probabilities corresponding to $S$ SAPs could be different. As we mentioned above, the occurrence probability of an SAP $p(s)$ is associated with the depth $d(s)$ of its representative leaf on the Huffman tree by the relation $p(s)=\left(\frac{1}{2}\right)^{d(s)}$. That is, the SAPs paired to leaves with larger depths will be activated less, and vice versa. As a result, we can design and optimize the pairing relation between $S$ SAPs to the $S$ leaves on a given Huffman tree so as to reduce the activation probabilities of those undesirable SAPs (lexicographic larger ones). In this way, the average BLER is expected to be reduced, and hence a coding gain can be harvested. Note that the coding gain achieved by the proposed optimization strategy using CSI at the transmitter will disappear if the given Huffman tree is a perfect binary tree (i.e., all interior nodes on the tree have two children and all resultant leaves are with the same depth \cite{2008data}).

In particular, given a Huffman tree, we rank all its leaves by their depths in ascending order as\footnote{One should note that it is possible and even common that for $s\neq t$, we have $d_{\langle s \rangle}(\sigma_s)=d_{\langle t \rangle}(\sigma_t)$, which indicates that both leaves are equally preferable and can be ranked arbitrarily. When given a perfect binary tree, we will have $d_{\langle 1 \rangle}(\sigma_1)= d_{\langle 2 \rangle}(\sigma_1)=\dots= d_{\langle s \rangle}(\sigma_s)$, and all leaves are equivalent. As a result, the ordering process is nullified and no performance gain can be attained in this special case. }
\begin{equation}\label{hdsakj2}
d_{\langle 1 \rangle}(\sigma_1)\leq d_{\langle 2 \rangle}(\sigma_2)\leq\dots\leq d_{\langle S \rangle}(\sigma_S),
\end{equation}
where $d_{\langle s \rangle}(\sigma_s)$ is the $s$th smallest depth and $\sigma_s$ is the index of the leaf with the $s$th smallest depth. 

\begin{figure}[!t]
\centering
\includegraphics[width=5.5in]{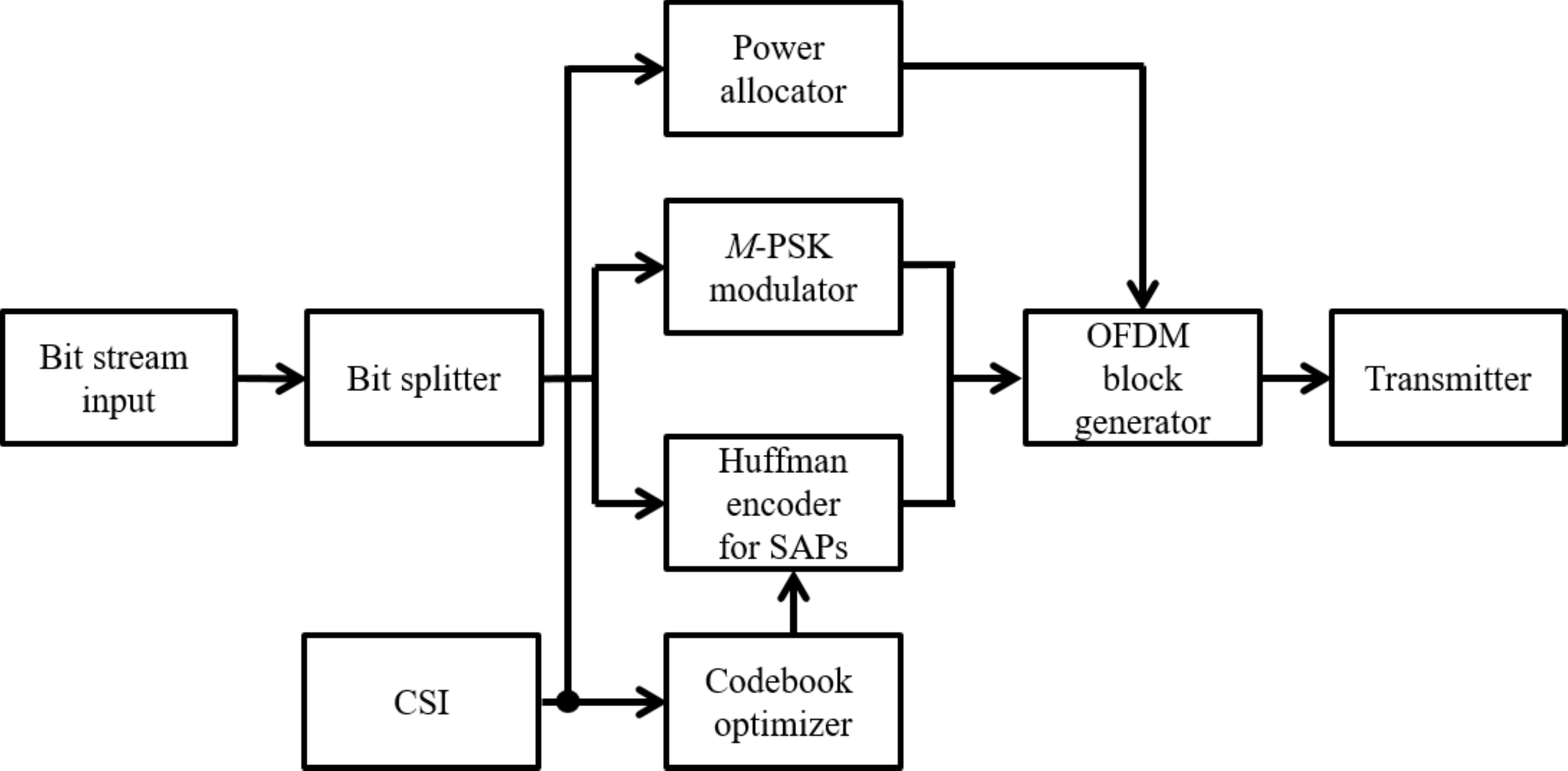}
\caption{System block diagram of the proposed EHC-OFDM-IM system.}
\label{sys}
\end{figure}

With the ranked relations presented in (\ref{2132131231555s}) and (\ref{hdsakj2}), we define the optimized bijective mapping relation between SAPs represented by their ASVs and leaves as\footnote{To enable the CSI-based mapping optimization, we assume that channels are slowly faded and the overhead caused by updating the optimized bijective mapping relation is negligible.}
\begin{equation}\label{fsjdhjbijec}
f_{\mathrm{opt}}:\mathbf{v}_s\leftrightarrows \sigma_s.
\end{equation}
By introducing the encoding procedure, we depict the system block diagram of the proposed EHC-OFDM-IM system in Fig. \ref{sys}. We present an example here to show how to implement the optimization strategy elaborated above.

\begin{example}
Assume $\{N,K\}=\{4,2\}$, i.e., four subcarriers in total and two of them are activated to form an SAP. Obviously, we have $S=\binom{4}{2}=6$ SAPs, which can be expressed by ASVs ranked by the lexicographic ordering principle as $\mathbf{v}_1=[0,0,1,1]^T$, $\mathbf{v}_2=[0,1,0,1]^T$, $\mathbf{v}_3=[0,1,1,0]^T$, $\mathbf{v}_4=[1,0,0,1]^T$, $\mathbf{v}_5=[1,0,1,0]^T$, and $\mathbf{v}_6=[1,1,0,0]^T$. Assuming the channel power gains of the four subcarriers are given by $G(1)=0.627$, $G(2)=0.884$, $G(3)=1.716$, and $G(4)=0.337$, we can then have the ranked channel power gains in ascending order as
\begin{equation}
G_{\left\langle 1\right\rangle}(4)<G_{\left\langle 2\right\rangle}(1)<G_{\left\langle 3\right\rangle}(2)<G_{\left\langle 4\right\rangle}(3).
\end{equation}
As ASVs are the collections of activation indicators of \textit{ordered} subcarriers, the six ASVs from $\mathbf{v}_1$ to $\mathbf{v}_6$ correspond to the subsets of active subcarriers $\{2,3\}$, $\{1,3\}$, $\{1,2\}$, $\{3,4\}$, $\{2,4\}$, and $\{1,4\}$, respectively. To provide an intuitive illustration of the full SAP ordering procedure, we pictorially demonstrate the given example in Fig. \ref{intactproc}.

\begin{figure*}[!t]
\centering
\includegraphics[width=6.0in]{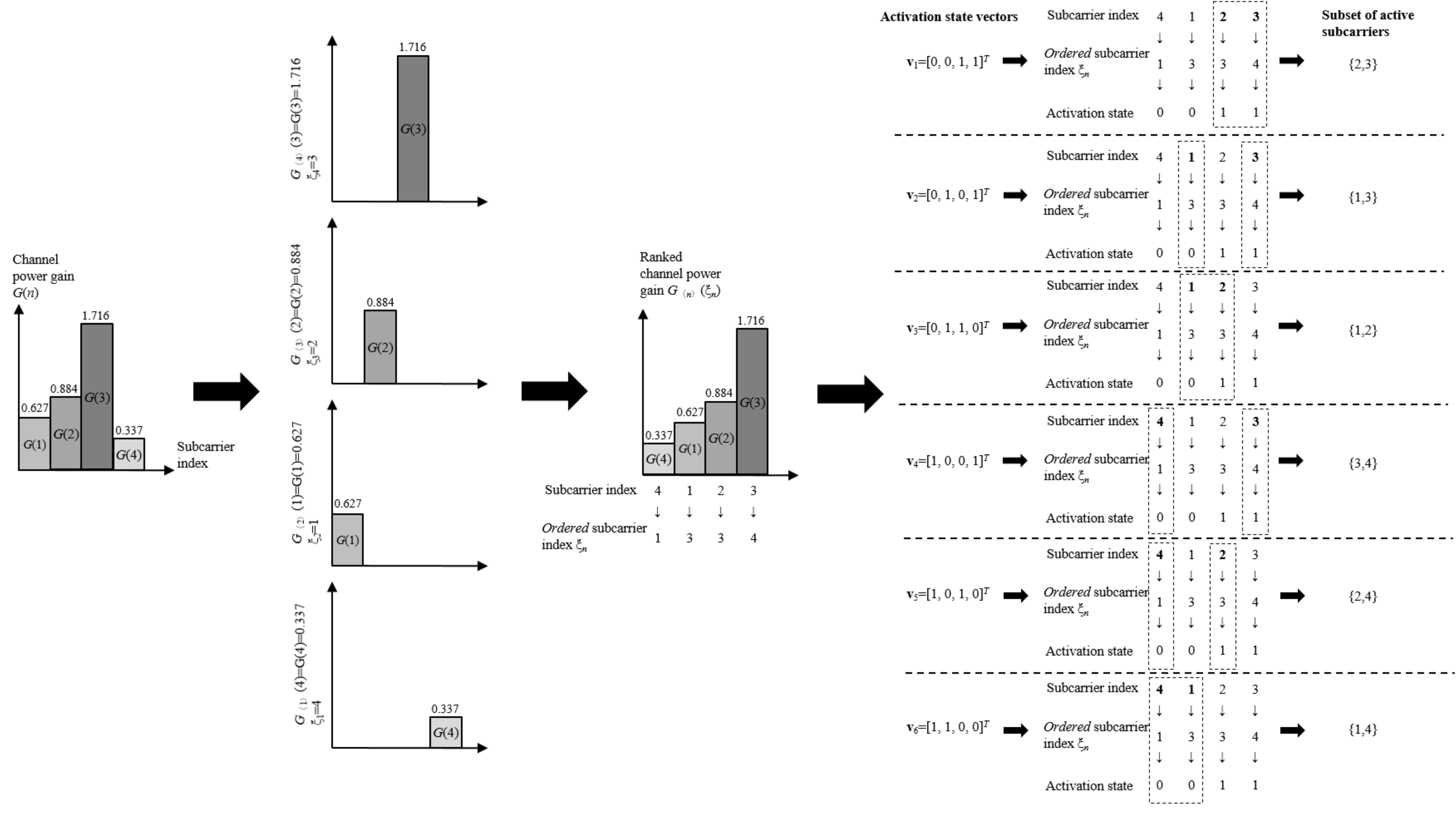}
\caption{SAP ordering procedure of the given example by the lexicographic ordering principle.}
\label{intactproc}
\end{figure*}

The Huffman tree with leaves numbered from left to right is shown in Fig. \ref{tree}.  Obviously, we have the ranked depths by
\begin{equation}
\underbrace{d_{\langle 1 \rangle}(5)= d_{\langle 2 \rangle}(6)}_{=2}<\underbrace{d_{\langle 3 \rangle}(1)=d_{\langle 4 \rangle}(2)= d_{\langle 5 \rangle}(3)= d_{\langle 6 \rangle}(4)}_{=3}.
\end{equation}
Subsequently, we resort to (\ref{fsjdhjbijec}) to yield the bijective mapping relation between the six SAPs and the six leaves in Table \ref{orderindasv}. By observing this table, we find that the two most preferable SAPs  represented by $\mathbf{v}_1$ and $\mathbf{v}_2$ are activated with higher probabilities than the other SAPs. By this optimized mapping relation, the resultant system is expected to provide a lower average BLER.

\begin{figure}[!t]
\centering
\includegraphics[width=5.0in]{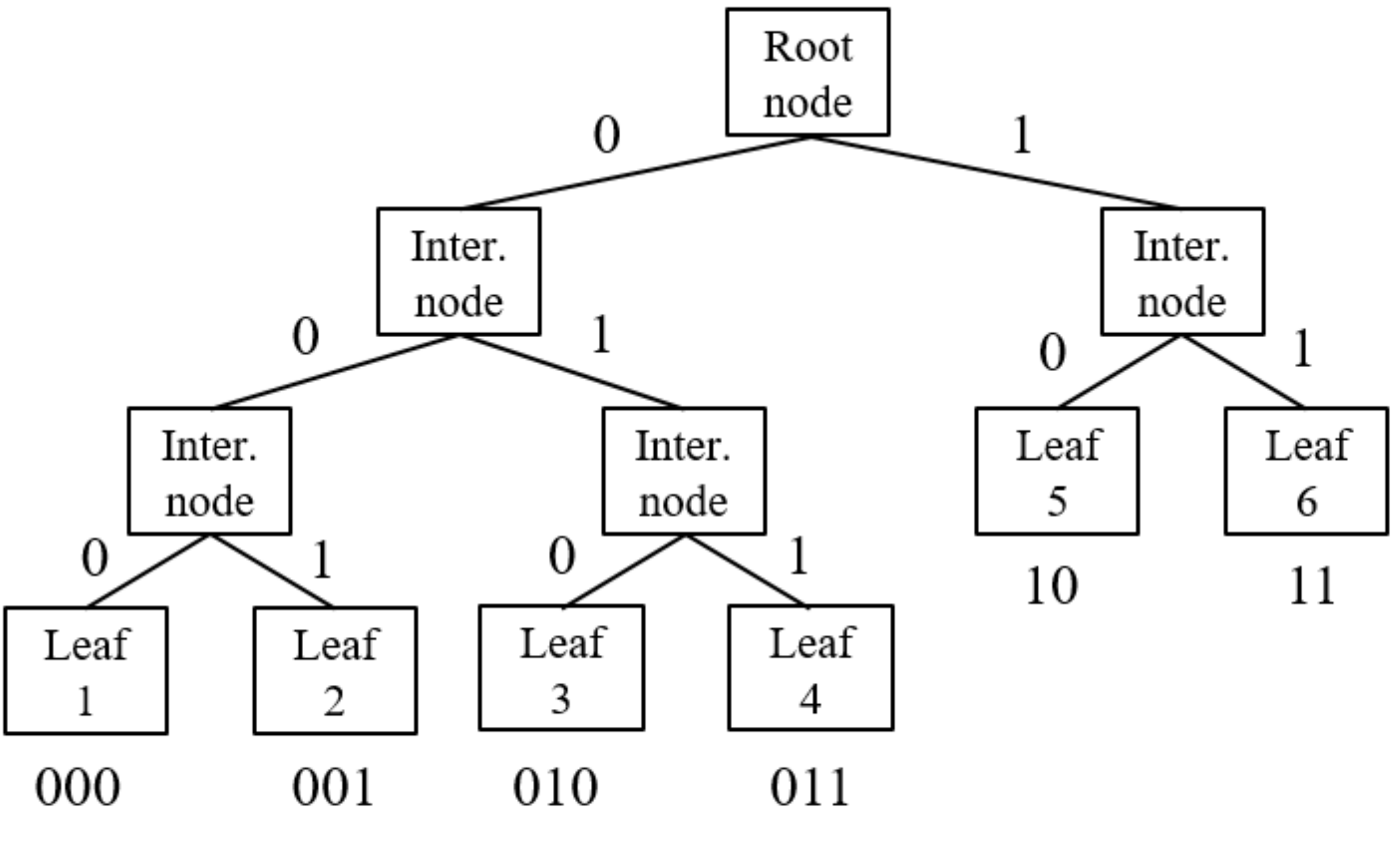}
\caption{An example of the Huffman tree with codewords attached.}
\label{tree}
\end{figure}

\begin{table}[!t]
\renewcommand{\arraystretch}{1.3}
\caption{An example of the bijective mapping relation between six SAPs and six leaves.}
\label{orderindasv}
\centering
\begin{tabular}{c|c|c|c|c|c}
\hline
ASV & \begin{tabular}{@{}c@{}}Active \\ subcarriers\end{tabular} & \begin{tabular}{@{}c@{}}Leaf's \\ depth\end{tabular} & \begin{tabular}{@{}c@{}}Leaf's \\ activation prob.\end{tabular}& \begin{tabular}{@{}c@{}}Leaf's \\ index\end{tabular}& \begin{tabular}{@{}c@{}}Code \\ words\end{tabular}\\
\hline\hline
$\mathbf{v}_1$ & $\{2,3\}$ & 2 & 1/4 & 5 & 10\\
$\mathbf{v}_2$ & $\{1,3\}$ & 2 & 1/4 & 6 & 11\\
$\mathbf{v}_3$ & $\{1,2\}$ & 3 & 1/8 & 1 & 000\\
$\mathbf{v}_4$ & $\{3,4\}$ & 3 & 1/8 & 2 & 001\\
$\mathbf{v}_5$ & $\{2,4\}$ & 3 & 1/8 & 3 & 010\\
$\mathbf{v}_6$ & $\{1,4\}$ & 3 & 1/8 & 4 & 011\\
\hline

\hline
\end{tabular}
\end{table}

\end{example}

\section{Error Performance Analysis}\label{epa}
For analytical simplicity, we temporarily assume that the transmit power allocated to each active subcarrier does not depend on CSI. Based on this regulation, we are able to derive a close-form and insightful expression of the average BLER. To do so, we first utilize the union bound to approximate (\ref{blerasjdh2}) by \cite{8640102}
\begin{equation}\label{dizokakek21s}
P_e(\mathbf{x}(s,\mathbf{m}))\leq \sum_{\hat{\mathbf{x}}(\hat{s},\hat{\mathbf{m}})\neq \mathbf{x}(s,\mathbf{m})}P_e\left(\mathbf{x}(s,\mathbf{m})\rightarrow\hat{\mathbf{x}}(\hat{s},\hat{\mathbf{m}})\right),
\end{equation}
where $P_e\left(\mathbf{x}(s,\mathbf{m})\rightarrow\hat{\mathbf{x}}(\hat{s},\hat{\mathbf{m}})\right)$ denotes  the pairwise error probability (PEP) that the actual transmitted OFDM block $\mathbf{x}(s,\mathbf{m})$ is  detected as $\hat{\mathbf{x}}(\hat{s},\hat{\mathbf{m}})$ by mistake. Further, PEP $P_e\left(\mathbf{x}(s,\mathbf{m})\rightarrow\hat{\mathbf{x}}(\hat{s},\hat{\mathbf{m}})\right)$ can be derived by averaging the conditional PEP over channel coefficients as
\begin{equation}
P_e\left(\mathbf{x}(s,\mathbf{m})\rightarrow\hat{\mathbf{x}}(\hat{s},\hat{\mathbf{m}})\right)=\underset{\mathbf{H}}{\mathbb{E}}\left\lbrace P_e\left(\mathbf{x}(s,\mathbf{m})\rightarrow\hat{\mathbf{x}}(\hat{s},\hat{\mathbf{m}})\vert \mathbf{H}\right)\right\rbrace.
\end{equation}
According to \cite{6587554}, when the ML detection scheme is in use, the conditional PEP can be determined in (\ref{dsakdjaksjd2keeq}), where $Q(\epsilon)=\frac{1}{\sqrt{2\pi}}\int_{\epsilon}^{\infty}\mathrm{exp}\left(-\frac{u^2}{2}\right)\mathrm{d}u$ is the Q-function; $(\mathrm{a})$ is approximated by an accurate exponential bound of the Q-function $Q(\epsilon)\approx\prod_{q=1}^{2}\rho_q\mathrm{exp}(-\eta_q\epsilon^2)$ proposed in \cite{1188428}; $\{\rho_1,\rho_2\}=\{1/12,1/4\}$ and $\{\eta_1,\eta_2\}=\{1/2,2/3\}$.

\begin{figure*}
\begin{equation}\label{dsakdjaksjd2keeq}\small
\begin{split}
P_e\left(\mathbf{x}(s,\mathbf{m})\rightarrow\hat{\mathbf{x}}(\hat{s},\hat{\mathbf{m}})\vert \mathbf{H}\right)&=Q\left(\sqrt{\frac{1}{N_0}\begin{Vmatrix}\mathbf{H}\left(\mathbf{\Omega}(s)\mathbf{x}(s,\mathbf{m})-\mathbf{\Omega}(\hat{s})\hat{\mathbf{x}}(\hat{s},\hat{\mathbf{m}})\right)\end{Vmatrix}_F^2}\right)\\
&=Q\left(\sqrt{\frac{1}{N_0}\sum_{n=1}^{N}G(n)|\sqrt{P_t(s,n)}x(s,m_n)-\sqrt{P_t(\hat{s},n)}\hat{x}(\hat{s},\hat{m}_n)|^2}\right)\\
&\overset{(\mathrm{a})}{\approx}\sum_{q=1}^{2}\rho_q \mathrm{exp}\left(-\frac{\eta_q}{N_0}\sum_{n=1}^{N}G(n)|\sqrt{P_t(s,n)}x(s,m_n)-\sqrt{P_t(\hat{s},n)}\hat{x}(\hat{s},\hat{m}_n)|^2\right)\\
&=\sum_{q=1}^{2}\rho_q \prod_{n=1}^{N}\mathrm{exp}\left(-\frac{\eta_q}{N_0}G(n)|\sqrt{P_t(s,n)}x(s,m_n)-\sqrt{P_t(\hat{s},n)}\hat{x}(\hat{s},\hat{m}_n)|^2\right)
\end{split}
\end{equation}
\hrule
\end{figure*}

To facilitate the following analysis and provide the simplest form of the final derivation, it is suggested to rearrange the form of the transmit OFDM block by the \textit{orders} of subcarriers instead of their \textit{indices}. Therefore, we resort to the concept of the \textit{permuted} OFDM block proposed in \cite{8519769} and have\footnote{One should note that the permuted OFDM block is simply utilized to facilitate the expression of analytical results and does not amend the actual transmission procedure at all.}
\begin{equation}
\mathbf{z}(s,\mathbf{m})=[z(s,m_{\xi_1}),z(s,m_{\xi_2}),\dots,z(s,m_{\xi_N})]^T\in\mathbb{C}^{N\times 1},
\end{equation}
which is built by replacing the `1's of an ASV with the data symbols in the corresponding $\mathbf{x}(s,\mathbf{m})$ in sequence. By establishing such an one-to-one mapping relation between $\mathbf{x}(s,\mathbf{m})$ and $\mathbf{z}(s,\mathbf{m})$, (\ref{dsakdjaksjd2keeq}) can be equivalently expressed as (\ref{lajio2es965}).

\begin{figure*}
\begin{equation}\label{lajio2es965}\small
\begin{split}
&P_e\left(\mathbf{x}(s,\mathbf{m})\rightarrow\hat{\mathbf{x}}(\hat{s},\hat{\mathbf{m}})\vert \mathbf{H}\right)\approx \sum_{q=1}^{2}\rho_q\prod_{n=1}^{N}\mathrm{exp}\left(-\frac{\eta_q}{N_0}G_{\langle n\rangle}(\xi_n)|\sqrt{P_t(s,\xi_n)}z(s,m_n)-\sqrt{P_t(\hat{s},\xi_n)}\hat{z}(\hat{s},\hat{m}_n)|^2\right)
\end{split}
\end{equation}
\hrule
\end{figure*}

Meanwhile, the PDF of the $n$th smallest channel power gain $G_{\langle n\rangle}(\xi_n)$ is determined by the rudiments of order statistics as \cite{david2004order}
\begin{equation}
\phi_{\langle n\rangle}(\epsilon)=\frac{N!(F_G(\epsilon))^{n-1}(1-F_G(\epsilon))^{N-n}f_G(\epsilon)}{(n-1)!(N-n)!},
\end{equation}
where $(\cdot)!$ denotes the factorial of the enclosed argument. Subsequently, we can remove the condition on $\mathbf{H}$ by averaging (\ref{lajio2es965}) over $\{G_{\langle n\rangle}(\xi_n)\}_{n=1}^{N}$ when the allocated power for active subcarriers is independent of $\mathbf{H}$, which yields (\ref{jidsaj2d1sa2}), where $(\mathrm{a})$ follows from by the independence among subcarriers and $\Gamma(\epsilon)=\int_{0}^{\infty}u^{\epsilon-1}\mathrm{exp}(-u)\mathrm{d}u$ is the complete gamma function. Substituting (\ref{jidsaj2d1sa2}) into (\ref{dizokakek21s}) gives the approximate BLER when $\mathbf{x}(s,\mathbf{m})$ is transmitted.

\begin{figure*}
\begin{equation}\label{jidsaj2d1sa2}\small
\begin{split}
P_e\left(\mathbf{x}(s,\mathbf{m})\rightarrow\hat{\mathbf{x}}(\hat{s},\hat{\mathbf{m}})\right)&=\underbrace{\int_{0}^{\infty}\int_{0}^{\infty}\dots\int_{0}^{\infty}}_{N}P_e\left(\mathbf{x}(s,\mathbf{m})\rightarrow\hat{\mathbf{x}}(\hat{s},\hat{\mathbf{m}})\vert \mathbf{H}\right)\left(\prod_{n=1}^{N}\phi_{\langle n\rangle}(G_{\langle n\rangle}(\xi_n))\right)\mathrm{d}G_{\langle 1\rangle}(\xi_1),\dots,\mathrm{d}G_{\langle N\rangle}(\xi_N)\\
&\overset{(\mathrm{a})}{=}\sum_{q=1}^{2}\rho_q\prod_{n=1}^{N}\frac{N!\Gamma\left(N-n+1+\frac{\eta_q}{N_0}|\sqrt{P_t(s,\xi_n)}z(s,m_{\xi_n})-\sqrt{P_t(\hat{s},\xi_n)}\hat{z}(\hat{s},\hat{m}_{\xi_n})|^2\right)}{(N-n)!\Gamma\left(N+1+\frac{\eta_q}{N_0}|\sqrt{P_t(s,\xi_n)}z(s,m_{\xi_n})-\sqrt{P_t(\hat{s},\xi_n)}\hat{z}(\hat{s},\hat{m}_{\xi_n})|^2\right)}
\end{split}
\end{equation}
\hrule
\end{figure*}

Finally, according to (\ref{pingjundasebls}), (\ref{fsjdhjbijec}) and the Huffman encoding rules, it is evident that there exist $M^K$ legitimate OFDM blocks adopting the same SAP. Hence, it is straightforward to explicitly express the average BLER as
\begin{equation}
\bar{P}_e=\frac{1}{M^K}\sum_{\mathbf{x}(s,\mathbf{m})\in\mathcal{X}}p(s)P_e(\mathbf{x}(s,\mathbf{m})).
\end{equation}

\section{Power Allocation and Error Performance Optimization}\label{pas}

\begin{figure}[!t]
\centering
\includegraphics[width=5.5in]{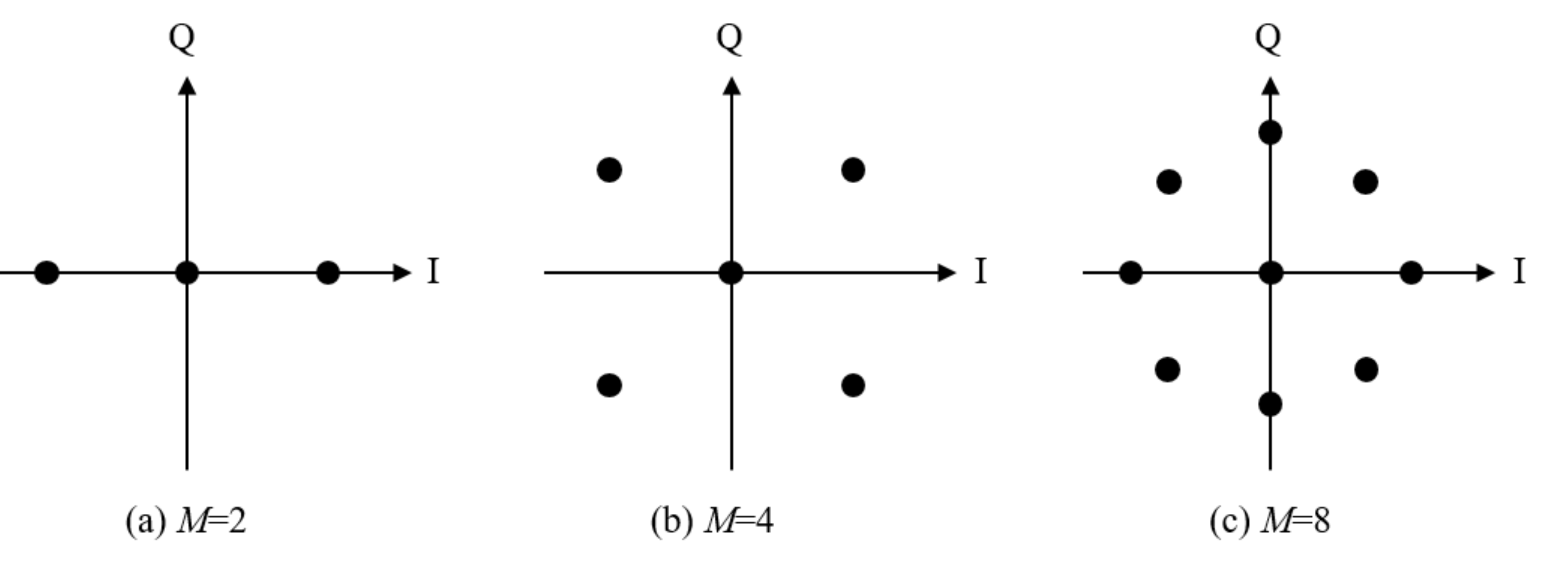}
\caption{Constellation diagrams for hypothetical quasi-OFDM-IM systems with $M=2$, $M=4$, and $M=8$ when $M$-PSK is used for data constellation modulation.}
\label{pa}
\end{figure}

\subsection{Suboptimal Power Allocation Scheme I}
When $\mathbf{H}$ is fixed and known at the transmitter, we can resort to the conditional average BLER as an optimization objective to formulate a power allocation problem to further enhance the error performance of EHC-OFDM-IM. Different from plain OFDM, by which all subcarriers are activated and $N$ symbols are transmitted, a subcarrier in EHC-OFDM-IM systems could be inactive. Also, because of the joint ML detection for symbols and SAP index (c.f. (\ref{sdakskjdk2dmldetec})), the power allocations for different SAPs are coupled. Both make the power allocation for EHC-OFDM-IM systems challenging. To enable the power allocation to optimize error performance, we make the following simplification. Specifically, from the perspective of transmitted symbols, deactivating a subcarrier is equivalent to transmitting a null on the subcarrier. In other words, we can reduce an EHC-OFDM-IM system utilizing $M$-PSK to a plain OFDM system utilizing a hypothetical $M+1$ constellation diagram on $K$ subcarriers, where the extra constellation point is the origin in the two-dimensional signal plane. We term this paradigm a quasi-OFDM-IM system. For clarity, we illustrate constellation diagrams for quasi-OFDM-IM systems with $M=2$, $M=4$, and $M=8$ in Fig. \ref{pa}. This simplification allows us to decouple the power allocation procedures among different SAPs.  

The BLER of the EHC-OFDM-IM system transmitting by the $s$th SAP can be approximated by the BLER of a hypothetical quasi-OFDM-IM system as\footnote{Note that, this approximation is applicable for both $M$-PSK, $M$-QAM and other data constellation modulation schemes, although we adopt $M$-PSK as an illustrative example in the following part. Changing data constellation modulation scheme will only result in a different $d_{\min}$.}
\begin{equation}\label{sdsajdhjkupskes5463362}
\Upsilon_e(s|\mathbf{H})=1-\prod_{i\in\mathcal{K}(s)}\left(1-Q\left(\sqrt{\frac{1}{N_0}P_t(s,i)G(i)d_{\min}^2}\right)\right),
\end{equation}
where $\mathcal{K}(s)$ is the set of $K$ active subcarriers in the $s$th SAP; $d_{\min}$ is the minimum Euclidean distance (MED) among $M+1$ constellation points, which can be determined by
\begin{equation}
d_{\min}=\min\left\lbrace 1,\sqrt{2\left(1-\cos\left({2\pi}/{M}\right)\right)}\right\rbrace,
\end{equation}
when $M$-PSK is used for data constellation modulation. In the high signal-to-noise (SNR) region, (\ref{sdsajdhjkupskes5463362}) can be simplified to \cite{6157252}
\begin{equation}
\Upsilon_e(s|\mathbf{H})\approx\tilde{\Upsilon}_e(s|\mathbf{H})=\sum_{i\in\mathcal{K}(s)}Q\left(\sqrt{\frac{1}{N_0}P_t(s,i)G(i)d_{\min}^2}\right).
\end{equation}
Now, we can formulate the power allocation problem for the $s$th SAP as (SP1):
\begin{equation*}
\begin{split}
(\mathrm{SP1})~~&\underset{\{P_t(s,n)\}}{\min}\left\lbrace\tilde{\Upsilon}_e(s|\mathbf{H})\right\rbrace\\
&~~~~\mathrm{s.t.} \sum_{i\in\mathcal{K}(s)}P_t(s,i)\leq\Psi\\
&~~~~~~~~~P_t(s,i)>0,~\forall~i \in\mathcal{K}(s)\\
&~~~~~~~~~P_t(s,i)=0,~\forall~i \not\in\mathcal{K}(s)
\end{split}
\end{equation*}
where $\Psi$ is the power budget allowed to be allocated to $K$ active subcarriers. Because the Q-function $Q(\cdot)$ is a monotone decreasing function, it is obvious that the formulated problem (SP1) is a quasi-convex optimization problem with respect to $\{P_t(s,n)\}$ (it can be rigorously proved by the positive second-order partial derivative of the objective function with respect to $\{P_t(s,n)\}$). As a consequence, we can construct a Lagrange function associated with the formulated problem (SP1) as follows:
\begin{equation}\label{lagdsjhequas}
\begin{split}
&\mathcal{L}_1(P_t(s,1),P_t(s,2),\dots,P_t(s,N),\lambda)\\
&~~~~=\tilde{\Upsilon}_e(s|\mathbf{H})-\lambda\left(\Psi-\sum_{i\in\mathcal{K}(s)}^{}P_t(s,i)\right),
\end{split}
\end{equation}
where $\lambda$ is the Lagrange multiplier. Subsequently, we resort to the Craig's expression of the Q-function and the Lebesgue's dominated convergence theorem to establish the Lagrange equation set as\footnote{The Lebesgue's dominated convergence theorem enables us to exchange the order of integration and differentiation operations in the constructed Lagrange function under certain conditions \cite{folland2013real}.}
\begin{equation}\label{dsajdh2jdsaiosbas}
\begin{cases}
\frac{\partial\mathcal{L}_1}{\partial \lambda}=-\left(\Psi-\underset{i\in\mathcal{K}(s)}{\sum}P_t(s,i)\right)=0\\
\frac{\partial\mathcal{L}_1}{\partial P_t(s,n)}=-\frac{1}{2\sqrt{\pi}P_t(s,n)}\sqrt{\frac{P_t(s,n)G(n)d_{\min}^2}{2N_0}}\\
~~~~~~~~~~~~\times\mathrm{exp}\left(-\frac{P_t(s,n)G(n)d_{\min}^2}{2N_0}\right)+\lambda=0,~\forall~n\in\mathcal{K}(s)
\end{cases}
\end{equation}
Solving (\ref{dsajdh2jdsaiosbas}) yields the power allocation scheme to (SP1) when the $s$th SAP is in use.

However, as solving (\ref{dsajdh2jdsaiosbas}) involves transcendental equations, to the best of authors' knowledge, there is no closed-form expression of the allocated power to each active subcarrier. Iterative methods can be applied to obtain numerical solutions to (SP1). We take the bisection method as an example to numerically obtain the solution due to its simplicity and robustness \cite{gomes2009implicit}. To do so, we first carry out a simple transformation of the second equation in (\ref{dsajdh2jdsaiosbas}) and obtain the expression of $P_t(s,t)$ as a function of $\lambda$:
\begin{equation}\label{ptsnkinversqeq}
P_t(s,n)=\frac{N_0}{G(n)d_{\min}^2}W\left(\frac{1}{8\pi}\left(\frac{G(n)d_{\min}^2}{\lambda N_0}\right)^2\right),
\end{equation}
where $W(\epsilon)$ is the Lambert W function giving the principal solution for $\varrho$ in $\epsilon=\varrho\mathrm{exp}(\varrho)$. We substitute (\ref{ptsnkinversqeq}) into the first equation in (\ref{dsajdh2jdsaiosbas}) and construct a function $\Theta(\lambda)$ of $\lambda$ as
\begin{equation}\label{dasdhj25541}
\Theta(\lambda)=\Psi-\frac{N_0}{d_{\min}^2}\sum_{i\in\mathcal{K}(s)}\frac{1}{G(i)}W\left(\frac{1}{8\pi}\left(\frac{G(i)d_{\min}^2}{\lambda N_0}\right)^2\right).
\end{equation}

To perform the bisection method and start halving in an iterative manner, we first need to initialize the root interval $[\lambda_{-},\lambda_{+}]$. However, because of the complication of the Lambert W function, the lower and upper limits $\lambda_{-}$, $\lambda_{+}$ that ensure $\Theta(\lambda_{-})<0$ and $\Theta(\lambda_{+})>0$ are not straightforward, which require elaborate design considering the trade-off between computational complexity and robustness. To determine these two limits, we apply heuristic algorithms as described in {Algorithm \ref{searchlower}} and {Algorithm \ref{searchupper}} to properly initialize  $\lambda_{-}$ and $\lambda_{+}$.  Subsequently, with the initial root interval $[\lambda_{-},\lambda_{+}]$ and a stipulated precision $\varepsilon$, the exact solution $\lambda^*$ can be approached by the numerical solution $\lambda^\&$ according to the bisection method as shown in {Algorithm \ref{dichotomysearch}}. Obviously, as long as we can find a $\lambda^{\&}$ so as to achieve $\Theta(\lambda^{\&})\approx 0$, we can substitute the value of $\lambda^{\&}$ back into (\ref{ptsnkinversqeq}) to determine the amount of transmit power allocated to the $n$th active subcarrier when the $s$th SAP is in use. By repeating the procedure $S$ times, the complete power allocation scheme for all $S$ SAPs can be obtained.

\begin{algorithm}[!t]
\caption{Algorithm to find out the lower limit $\lambda_{-}$ to start the bisection method based approaching procedure.} 
\label{searchlower} 
\begin{algorithmic}[1] 
\STATE \textbf{BEGIN}
	\STATE Input: $N_0$, $\mathbf{H}$, $\mathcal{K}(s)$, and $M$;
	\STATE $j\leftarrow 0$;
	\STATE $\lambda_{-}\leftarrow \mathrm{exp}(j)$;
	\STATE Calculate $\Theta(\lambda_{-})$ by (\ref{dasdhj25541});
	\WHILE{$\Theta(\lambda_{-})>0$}
	\STATE $j--$;
	\STATE $\lambda_{-}\leftarrow \mathrm{exp}(j)$;
	\ENDWHILE
	\RETURN $\lambda_-$;
\STATE \textbf{END}
\end{algorithmic}
\end{algorithm}

\begin{algorithm}[!t]
\caption{Algorithm to find out the upper limit $\lambda_{+}$ to start the bisection method based approaching procedure.} 
\label{searchupper} 
\begin{algorithmic}[1] 
\STATE \textbf{BEGIN}
	\STATE Input: $N_0$, $\mathbf{H}$, $\mathcal{K}(s)$, and $M$;
	\STATE $j\leftarrow 0$;
	\STATE $\lambda_{+}\leftarrow \mathrm{exp}(j)$;
	\STATE Calculate $\Theta(\lambda_{+})$ by (\ref{dasdhj25541});
	\WHILE{$\Theta(\lambda_{+})<0$}
	\STATE $j++$;
	\STATE $\lambda_{+}\leftarrow \mathrm{exp}(j)$;
	\ENDWHILE
	\RETURN $\lambda_+$;
\STATE \textbf{END}
\end{algorithmic}
\end{algorithm}

\begin{algorithm}[!t]
\caption{Bisection method based procedure to approximate $\lambda^*$ with $\lambda^\&$ by a stipulated precision $\varepsilon$.} 
\label{dichotomysearch} 
\begin{algorithmic}[1] 
\STATE \textbf{BEGIN}
	\STATE Input: $\lambda_{-}$, $\lambda_{+}$, $\varepsilon$, $N_0$, $\mathbf{H}$, $\mathcal{K}(s)$, and $M$;
	\STATE $\lambda_{\mathsf{m}}\leftarrow(\lambda_{-}+\lambda_{+})/2$
	\WHILE{$\lambda_{\mathsf{m}}-\lambda_{-}>\varepsilon$}
		\STATE Calculate $\Theta(\lambda_{-})$ by (\ref{dasdhj25541});
		\STATE Calculate $\Theta(\lambda_{\mathsf{m}})$ by (\ref{dasdhj25541});
		\IF{$\Theta(\lambda_-)\Theta(\lambda_{\mathsf{m}})<0$}
			\STATE $\lambda_+\leftarrow\lambda_{\mathsf{m}}$;
		\ELSE
			\STATE $\lambda_-\leftarrow\lambda_{\mathsf{m}}$;
		\ENDIF
		\STATE $\lambda_{\mathsf{m}}\leftarrow(\lambda_{-}+\lambda_{+})/2$

	\ENDWHILE
	\STATE $\lambda^\&\leftarrow\lambda_{\mathsf{m}}$;
	\RETURN $\lambda^\&$;
\STATE \textbf{END}
\end{algorithmic}
\end{algorithm}

\subsection{Suboptimal Power Allocation Scheme II}\label{dsahdkj2hjdsreducpowsec}
Because numerically solving (SP1) would render high computational complexity and a great amount of computational time, especially when the number of active subcarriers $K$ is large, this could impair the practicality of power allocation for EHC-OFDM-IM in realistic scenarios where channel states vary rapidly. Therefore, to enhance the practicality, we are inspired by the exponential approximation of the Q-function $Q(\epsilon)\approx\prod_{q=1}^{2}\rho_q\mathrm{exp}(-\eta_q\epsilon^2)$ and reformulate another much simpler suboptimal power allocation problem (SP2) for a large power budget infra:
\begin{equation*}
\begin{split}
(\mathrm{SP2})~~&\underset{\{P_t(s,n)\}}{\min}\left\lbrace \sum_{i\in\mathcal{K}(s)}\mathrm{exp}\left(-\frac{1}{N_0}P_t(s,i)G(i)d_{\min}^2\right)\right\rbrace\\
&~~~~\mathrm{s.t.} \sum_{i\in\mathcal{K}(s)}P_t(s,i)\leq\Psi\\
&~~~~~~~~~P_t(s,i)>0,~\forall~n \in\mathcal{K}(s)\\
&~~~~~~~~~P_t(s,i)=0,~\forall~n \not\in\mathcal{K}(s)
\end{split}
\end{equation*}
which is also a convex optimization problem with respect to $\{P_t(s,n)\}$ (this can also be easily proved by the positive second-order partial derivative of the objective function with respect to $\{P_t(s,n)\}$). Analogously, the Lagrange function associated with the reformulated problem (SP2) is given by
\begin{equation}\label{lagdsjhequas222222}
\begin{split}
&\mathcal{L}_2(P_t(s,1),P_t(s,2),\dots,P_t(s,N),\lambda)\\
&~~~~=\sum_{i\in\mathcal{K}(s)}\mathrm{exp}\left(-\frac{1}{N_0}P_t(s,i)G(i)d_{\min}^2\right)\\
&~~~~~~~~~~~~~~~~-\lambda\left(\Psi-\sum_{i\in\mathcal{K}(s)}^{}P_t(s,i)\right).
\end{split}
\end{equation}
We can construct the Lagrange equation set for (\ref{lagdsjhequas222222}) to be
\begin{equation}\label{dsajdh2jdsaiosbas222222}
\begin{cases}
\frac{\partial\mathcal{L}_2}{\partial \lambda}=-\left(\Psi-\underset{i\in\mathcal{K}(s)}{\sum}P_t(s,i)\right)=0\\
\frac{\partial\mathcal{L}_2}{\partial P_t(s,n)}\\
=-\frac{G(n)d_{\min}^2}{N_0}\mathrm{exp}\left(-\frac{P_t(s,n)G(n)d_{\min}^2}{N_0}\right)+\lambda=0,~\forall~n\in\mathcal{K}(s)
\end{cases}
\end{equation}
For an arbitrary active subcarrier $n$, $\forall~n\in\mathcal{K}(s)$, according to $-\frac{G(n)d_{\min}^2}{N_0}\mathrm{exp}\left(-\frac{P_t(s,n)G(n)d_{\min}^2}{N_0}\right)+\lambda=0$, we can easily express $P_t(s,n)$ as a function of $\lambda$ by
\begin{equation}\label{dsakdj2kisptsn}
P_t(s,n)=-\frac{N_0}{G(n)d_{\min}^2}\log\left(\frac{\lambda N_0}{G(n)d_{\min}^2}\right).
\end{equation}
Substituting (\ref{dsakdj2kisptsn}) for all active subcarriers, $\forall~n\in\mathcal{K}(s)$, into $\left(\Psi-\underset{i\in\mathcal{K}(s)}{\sum}P_t(s,i)\right)=0$ and transforming the resulted expression yields
\begin{equation}
\Psi=-\frac{N_0}{d_{\min}^2}\sum_{i\in\mathcal{K}(s)}\frac{1}{G(i)}\log\left(\frac{\lambda N_0}{G(i)d_{\min}^2}\right).
\end{equation}
By the fundamental properties of logarithm, we obtain the solution of $\lambda$ to be
\begin{equation}\label{dsadhj2poduwwqqqq}
\lambda=\mathrm{exp}\left(-\frac{\Psi d_{\min}^2+N_0\underset{i\in\mathcal{K}(s)}{\sum}\frac{1}{G(i)}\log\left(\frac{N_0}{G(i)d_{\min}^2}\right)}{N_0\underset{i\in\mathcal{K}(s)}{\sum}\frac{1}{G(i)}}\right).
\end{equation}
Substituting (\ref{dsadhj2poduwwqqqq}) back into (\ref{dsakdj2kisptsn}) gives the solutions of $\{P_t(s,n)\}$, the amount of power allocated to active subcarriers. By repeating the procedure $S$ times, the complete power allocation scheme produced by the simplified optimization objective for all $S$ SAPs can be obtained.

\subsection{Tailored Classic Power Allocation Schemes}\label{opassstsc}
In order to provide comparison benchmarks, we also borrow two classic power allocation schemes from plain OFDM and tailor them for EHC-OFDM-IM, which are the uniform power allocation and the equalization power allocation. The power allocation outcomes for the $s$th SAP produced by the two tailored classic power allocation schemes are given by \cite{54342,1097882}
\begin{equation}
P_t(s,n)=\begin{cases}
\Psi/K,~~~~\forall~n\in\mathcal{K}(s)\\
0,~~~~~~~~~\forall~n\not\in\mathcal{K}(s)
\end{cases}
\end{equation}
and
\begin{equation}
P_t(s,n)=\begin{cases}
\left.\Psi\middle / \left(G(n)\underset{i\in\mathcal{K}(s)}{\sum}\frac{1}{G(i)}\right)\right.,~~~~\forall~n\in\mathcal{K}(s)\\
0,~~~~~~~~~~~~~~~~~~~~~~~~~~~~~~~~\forall~n\not\in\mathcal{K}(s)
\end{cases}
\end{equation}
By repeating the above allocation procedure $S$ times, the complete power allocation schemes for all $S$ SAPs by the tailored uniform and equalization power allocation criteria can be obtained.

\subsection{Analysis of Computational Complexity}
For comparison, we analyze the computational complexities of all depicted power allocation schemes in this subsection. The proposed sub-optimal power allocation scheme I needs to find $\lambda^{\&}$ by using the bisection method in {Algorithm \ref{dichotomysearch}}. Provided that {Algorithm \ref{dichotomysearch}} takes $N_{\mathrm{iter}}$ to converge, the function $\Theta(\lambda)$ needs to be calculated by $N_{\mathrm{iter}}$ times. For each calculation of $\Theta(\lambda)$, we need to compute the Lambert $W$-function $W(\epsilon)$ $K$ times. Since the Lambert $W$-function can be expanded by infinite series as $W(\epsilon)=\sum_{n=1}^{\infty}\frac{(-n)^{n-1}}{n!}\epsilon^n$ \cite{corless1997sequence}, the computational complexity for calculating $W(\epsilon)$ can be expressed by the Bachmann-Landau notation as $\mathcal{O}(q)$ if the $q$th order approximation is adopted. Note that, because the complexities vis-\`{a}-vis {Algorithm \ref{searchlower}} and {Algorithm \ref{searchupper}} and the complexity caused by using $\lambda^{\&}$ to obtain $\{P_t(s,i)\}$  are much smaller than that of {Algorithm \ref{dichotomysearch}}, we thereby omit them for simplicity in the analysis. Summarizing all above, the complexity of the sub-optimal power allocation scheme I can be expressed as $\mathcal{O}(N_{\mathrm{iter}}Kq)$. 

On the contrary, the proposed sub-optimal power allocation II neither needs iterations nor the computation of $W(\epsilon)$. Therefore, its complexity is only of the order $\mathcal{O}(K)$, which proves its simplicity compared to the proposed sub-optimal power allocation scheme I. 

For the two tailored classic power allocation schemes, the uniform power allocation requires the computational complexity of the order $\mathcal{O}(1)$ and the equalization power allocation requires the computational complexity of the order $\mathcal{O}(K)$. For clarity, we compare these four power allocation schemes by listing their complexities and requirements of CSI in Table \ref{comparelistpowalloc}.

\begin{table}
\centering
\caption{Comparisons among different power allocation schemes for EHC-OFDM-IM systems.}
\label{comparelistpowalloc}
\begin{tabular}{c|c|c}
\hline
Schemes&Complexity&CSI requirement\\
\hline
\hline
SPA I& $\mathcal{O}(N_{\mathrm{iter}}Kq)$&YES\\
\hline
SPA II& $\mathcal{O}(K)$&YES\\
\hline
UPA& $\mathcal{O}(1)$&NO\\
\hline
EPA& $\mathcal{O}(K)$&YES\\
\hline
\end{tabular}
\end{table}

\section{Numerical Results}\label{nr}
\begin{figure*}[!t]
\centering
\includegraphics[width=6.0in]{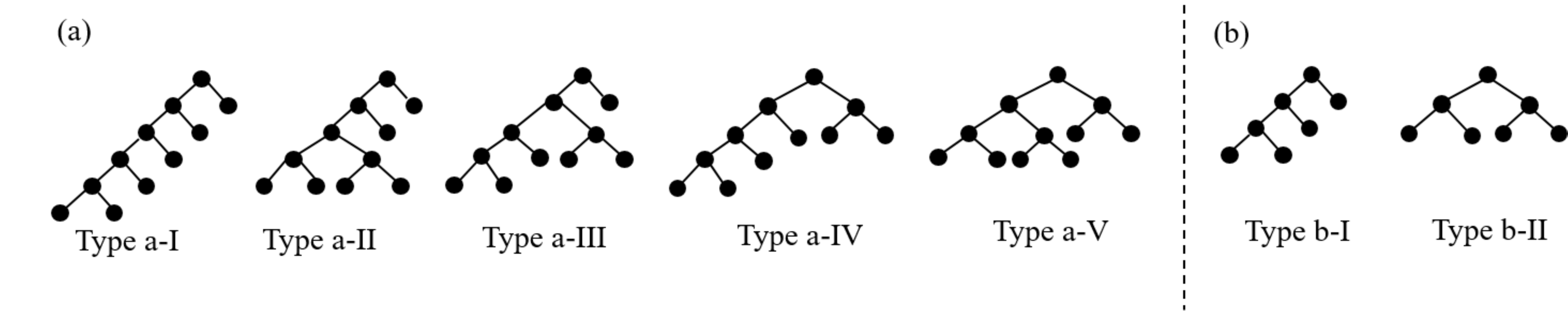}
\caption{Unique architectures of the Huffman trees for simulations: (a) Five types for  $\{N,K\}=\{4,2\}$ ($S=6$); (b) Two types for $\{N,K\}=\{4,3\}$ ($S=4$).}
\label{huffmantreeillu}
\end{figure*}
\subsection{Verification of Error Performance Analysis}

To be general, we select two typical setups with $\{N,K\}=\{4,2\}$ ($S=6$) and $\{N,K\}=\{4,3\}$ ($S=4$) to simulate in this section. For the case of $\{N,K\}=\{4,2\}$, there are five \textit{unique} architectures of the Huffman trees as shown in Fig. \ref{huffmantreeillu}(a). For the case of $\{N,K\}=\{4,3\}$, there are two unique architectures of the Huffman trees as shown in\footnote{The number of all possible architectures of full binary trees for the case with $S$ leaves is the Catalan number $C(S)$, which increases exponentially when $S$ increases and makes analysis difficult. However, as the tree architectures with symmetry can be regarded to be equivalent for error performance analysis, we can exclude them for simplicity, but without loss of generality. The approach to select unique tree architectures for an arbitrary number of SAPs has been detailed in \cite{8704951}.} Fig. \ref{huffmantreeillu}(b). We further normalize $\mu=1$, $N_0=1$ and utilize binary PSK (BPSK, $M=2$) for data constellation modulation on $K$ active subcarriers. To verify the error performance analysis when power allocation is independent of channel states, we tentatively adopt the uniform power allocation scheme that regulates $P_t(s,n)=\Psi/K$ for active subcarriers and is zero otherwise. We also adopt LCD-OFDM-IM proposed in \cite{8519769}, classic OFDM-IM proposed in \cite{6587554}, classic Huffman coded OFDM-IM proposed in \cite{7370760}, and plain OFDM as error performance comparison benchmarks.

\begin{figure*}[!t]
    \centering
    \begin{subfigure}[t]{0.5\textwidth}
        \centering
        \includegraphics[width=3.5in]{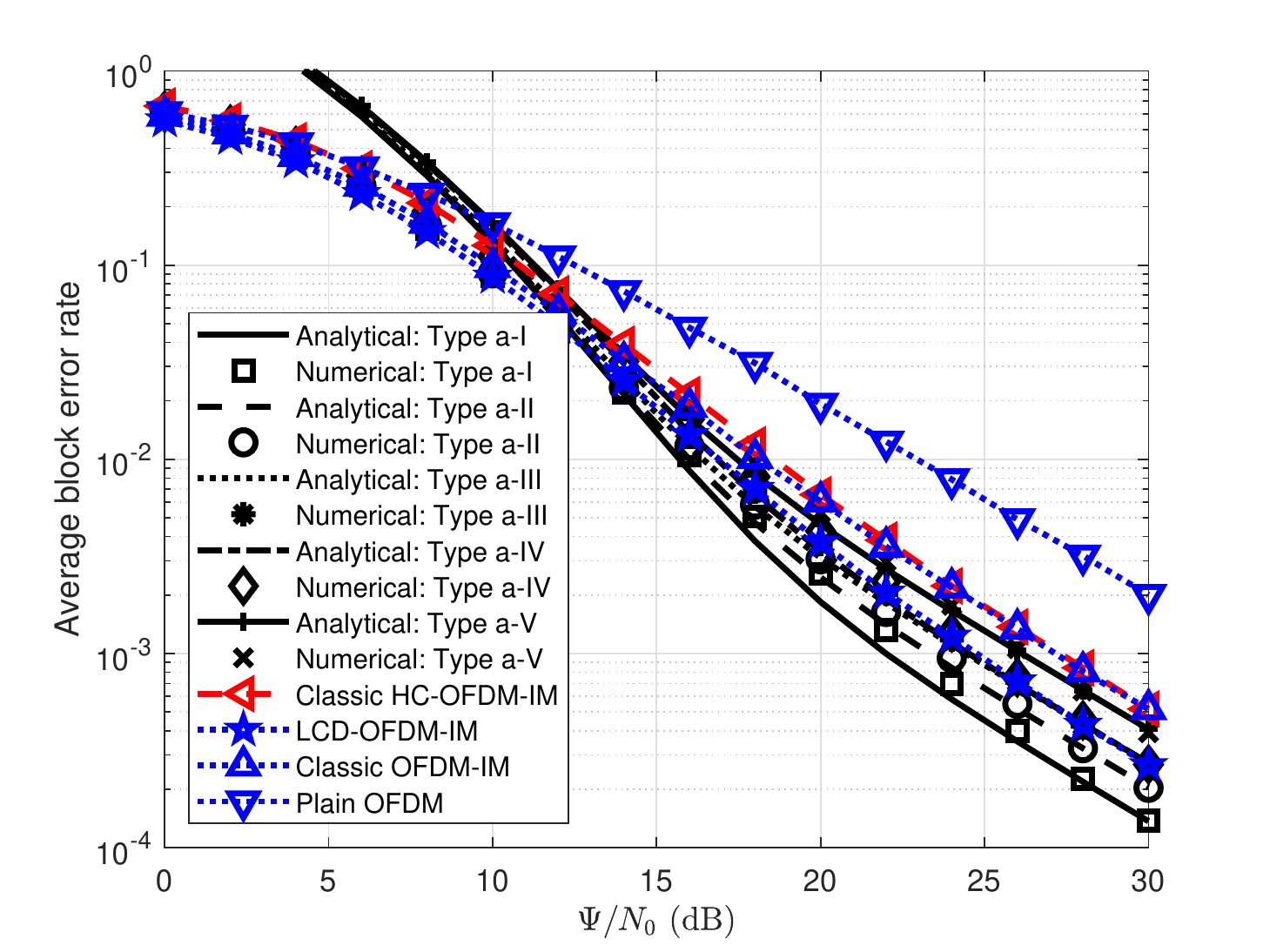}
        \caption{$\{N,K\}=\{4,2\}$}
    \end{subfigure}%
~
    \begin{subfigure}[t]{0.5\textwidth}
        \centering
        \includegraphics[width=3.5in]{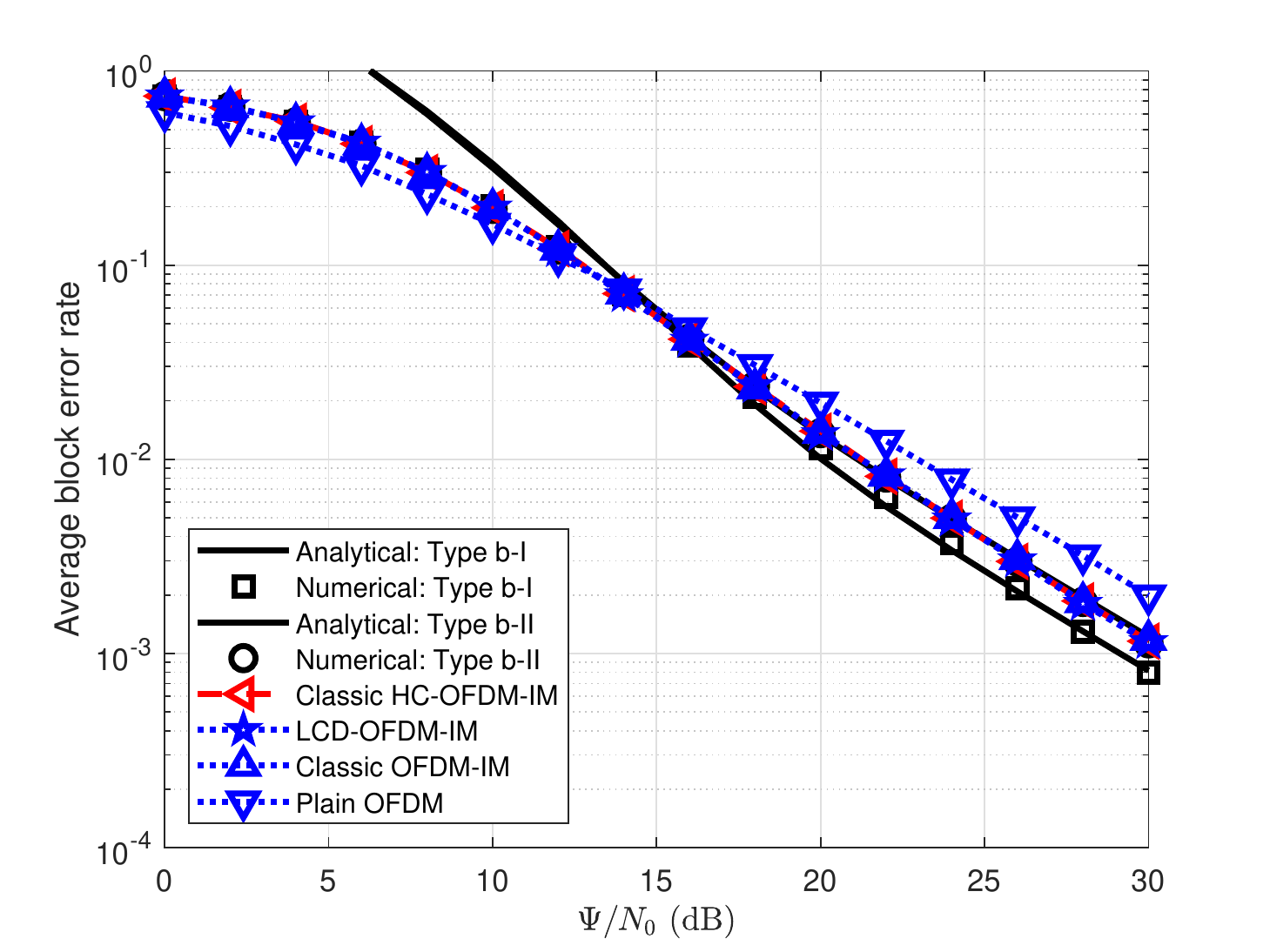}
        \caption{$\{N,K\}=\{4,3\}$}
    \end{subfigure}
    \caption{Average BLER vs. $\Psi/N_0$ with different tree architectures and different numbers of active subcarriers $K$, given $N=4$ and $M=2$.}
    \label{BLER_sim}
\end{figure*}

The analytical and numerical results regarding average BLER for the cases $\{N,K\}=\{4,2\}$ ($S=6$) and $\{N,K\}=\{4,3\}$ ($S=4$) are illustrated in Fig. \ref{BLER_sim}. From this figure, the effectiveness of the analysis presented in Section \ref{epa} is substantiated, as the analytical curves well approach the numerical curves at high SNR. The analytical errors existing in the low SNR region are mainly caused by the union bound (c.f. (\ref{dizokakek21s})), while the analytical errors existing in the high SNR region are mainly due to the exponential approximation of the Q-function (c.f. (\ref{dsakdjaksjd2keeq})). Meanwhile, by observing the numerical results presented in Fig. \ref{BLER_sim}, one can also gain insight into how the architecture of the Huffman tree affects the error performance. It is clear that the depth of the Huffman tree has a significant impact on the error performance when the SAP-to-leaf mapping relation is optimized based on CSI. In particular, a Huffman tree with a larger depth produces a lower average BLER. Therefore, Huffman trees a-I and b-I (with the largest depths) provide the lowest average BLER for EHC-OFDM-IM systems with $\{N,K\}=\{4,2\}$ and $\{N,K\}=\{4,3\}$, respectively.

\begin{figure}[!t]
\centering
\includegraphics[width=3.5in]{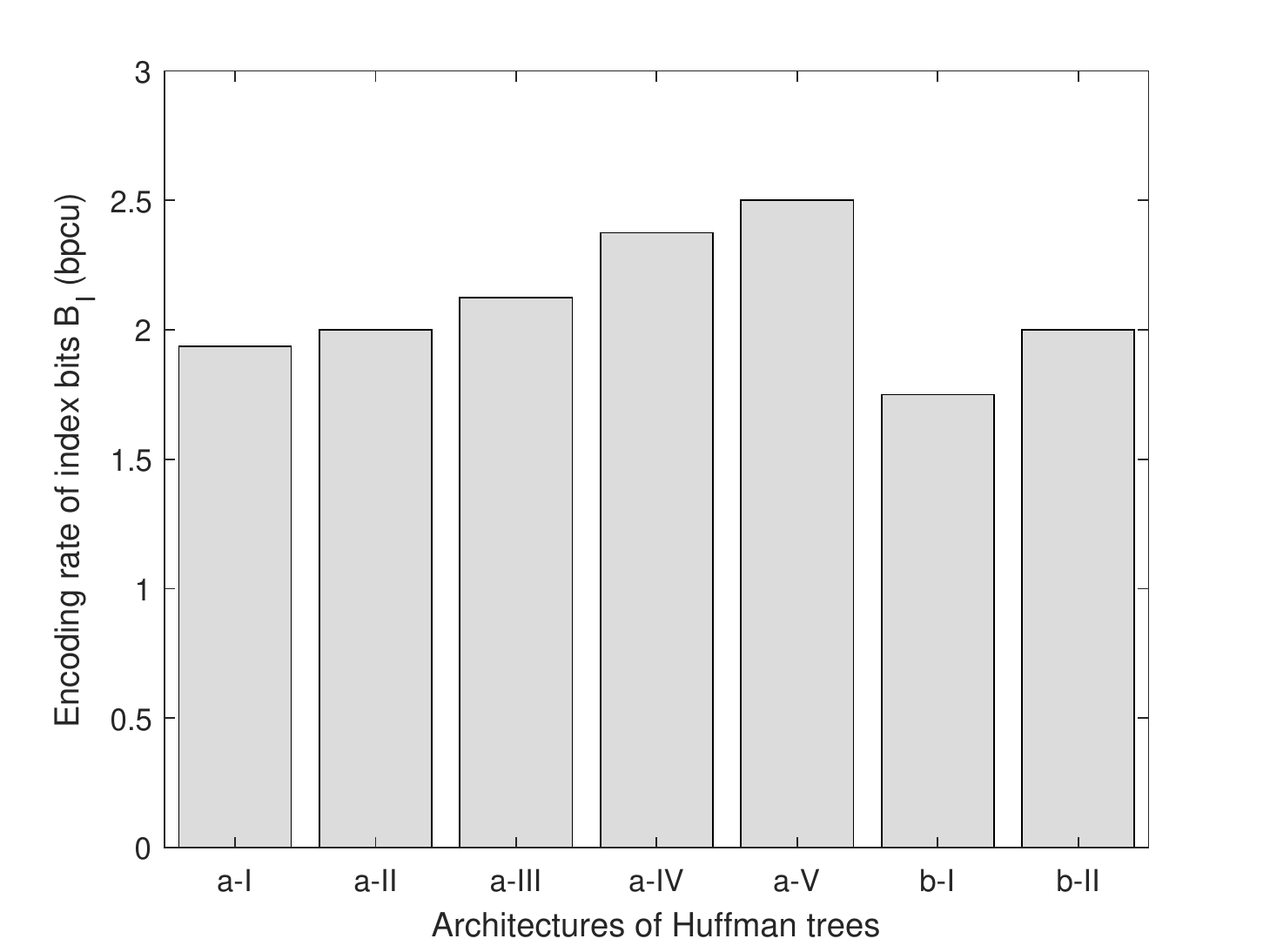}
\caption{Encoding rates of index bits associated with different tree architectures.}
\label{rate}
\end{figure}

On the other hand, the above observation does not imply that the Huffman tree with the largest depth is necessarily preferable. In fact, there is a trade-off in the architecture of the Huffman tree between error performance and data transmission rate. To clearly illustrate this trade-off, we plot the encoding rates of index bits associated with different tree architectures for EHC-OFDM-IM systems with $\{N,K\}=\{4,2\}$ and $\{N,K\}=\{4,3\}$ in Fig. \ref{rate}. From this figure, we can see that for the case of $\{N,K\}=\{4,2\}$, the Huffman tree a-I that produces the best error performance has the lowest encoding rate of index bits, while the Huffman tree a-V corresponding to the highest average BLER yields the highest encoding rate of index bits. The same trade-off can also be observed between Huffman trees b-I and b-II. To summarize, one should take both error performance and required transmission rate into consideration when choosing the architecture of the Huffman tree for EHC-OFDM-IM systems. As the optimization of the tree architecture is out of the scope of this paper, we omit in-depth discussion here.

In addition, compared to LCD-OFDM-IM and classic OFDM-IM, it is verified that one can adjust the tree architecture of EHC-OFDM-IM to achieve better error performance. However, if the tree architecture is improperly designed, EHC-OFDM-IM could even lead to a higher average BLER than LCD-OFDM-IM (refer to the cases of a-III and a-V). All cases of EHC-OFDM-IM outperform the cases of classic OFDM-IM and plain OFDM without IM. 

Furthermore, by the comparison between EHC-OFDM-IM and classic Huffman coded OFDM-IM, we can observe  that the proposed EHC-FODM-IM scheme is capable of harvesting a coding gain and outperforms the classic Huffman coded OFDM-IM scheme when the tree architecture is not perfect. On the other hand, utilizing a perfect tree architecture, i.e., tree architecture b-II, the coding gain brought by EHC-OFDM-IM vanishes, which aligns with our expectation, because all leaves in this case have the same depth and can be regarded as equivalent.

\begin{figure*}[!t]
    \centering
    \begin{subfigure}[t]{0.5\textwidth}
        \centering
        \includegraphics[width=3.5in]{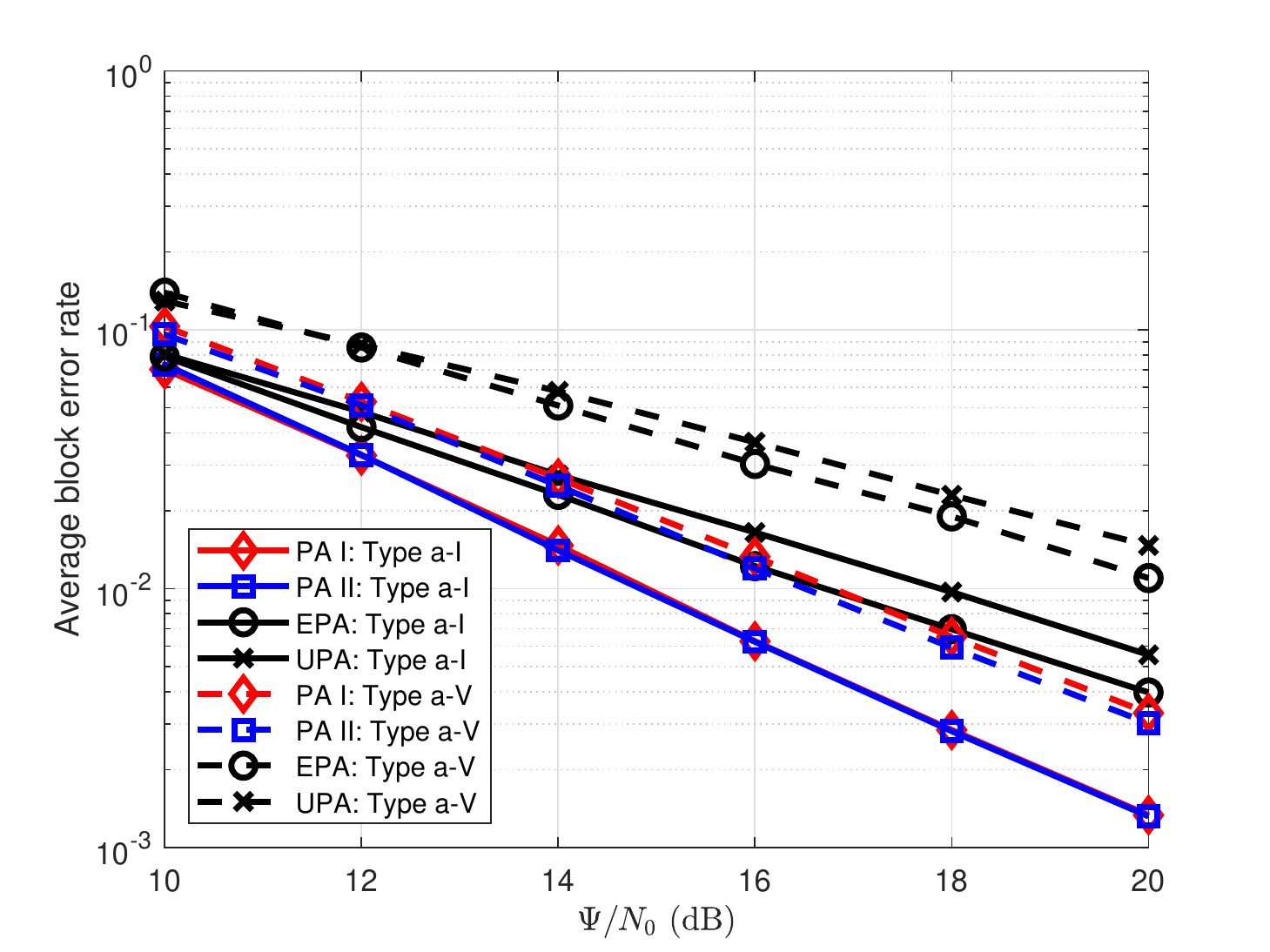}
        \caption{$\{N,K\}=\{4,2\}$}
    \end{subfigure}%
~
    \begin{subfigure}[t]{0.5\textwidth}
        \centering
        \includegraphics[width=3.5in]{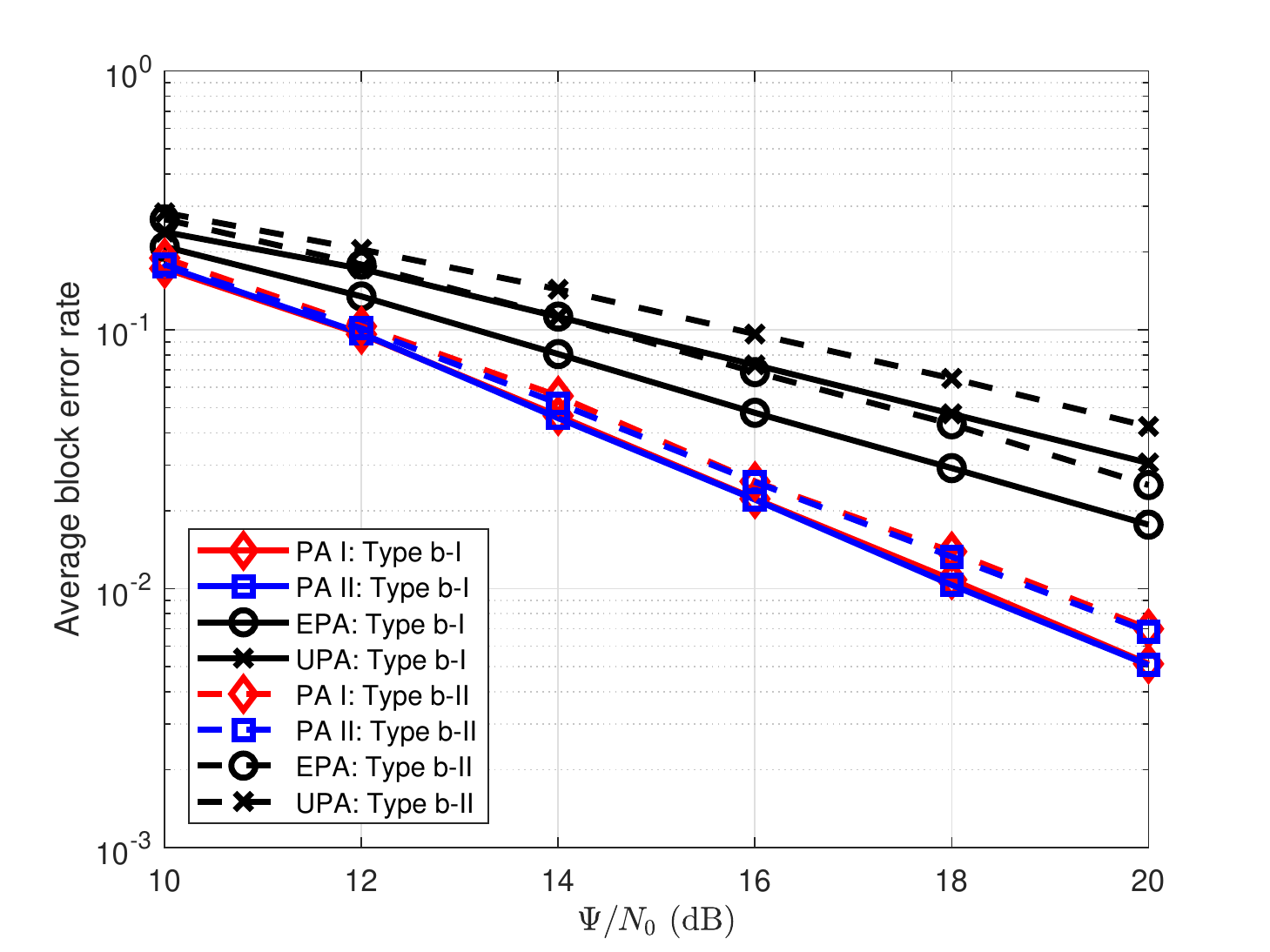}
        \caption{$\{N,K\}=\{4,3\}$}
    \end{subfigure}
    \caption{Average BLER comparisons among different power allocation schemes with different tree architectures and different numbers of active subcarriers $K$, given $N=4$ and $M=2$.}
    \label{PAsim}
\end{figure*}

\begin{figure*}[!t]
    \centering
    \begin{subfigure}[t]{0.5\textwidth}
        \centering
        \includegraphics[width=3.5in]{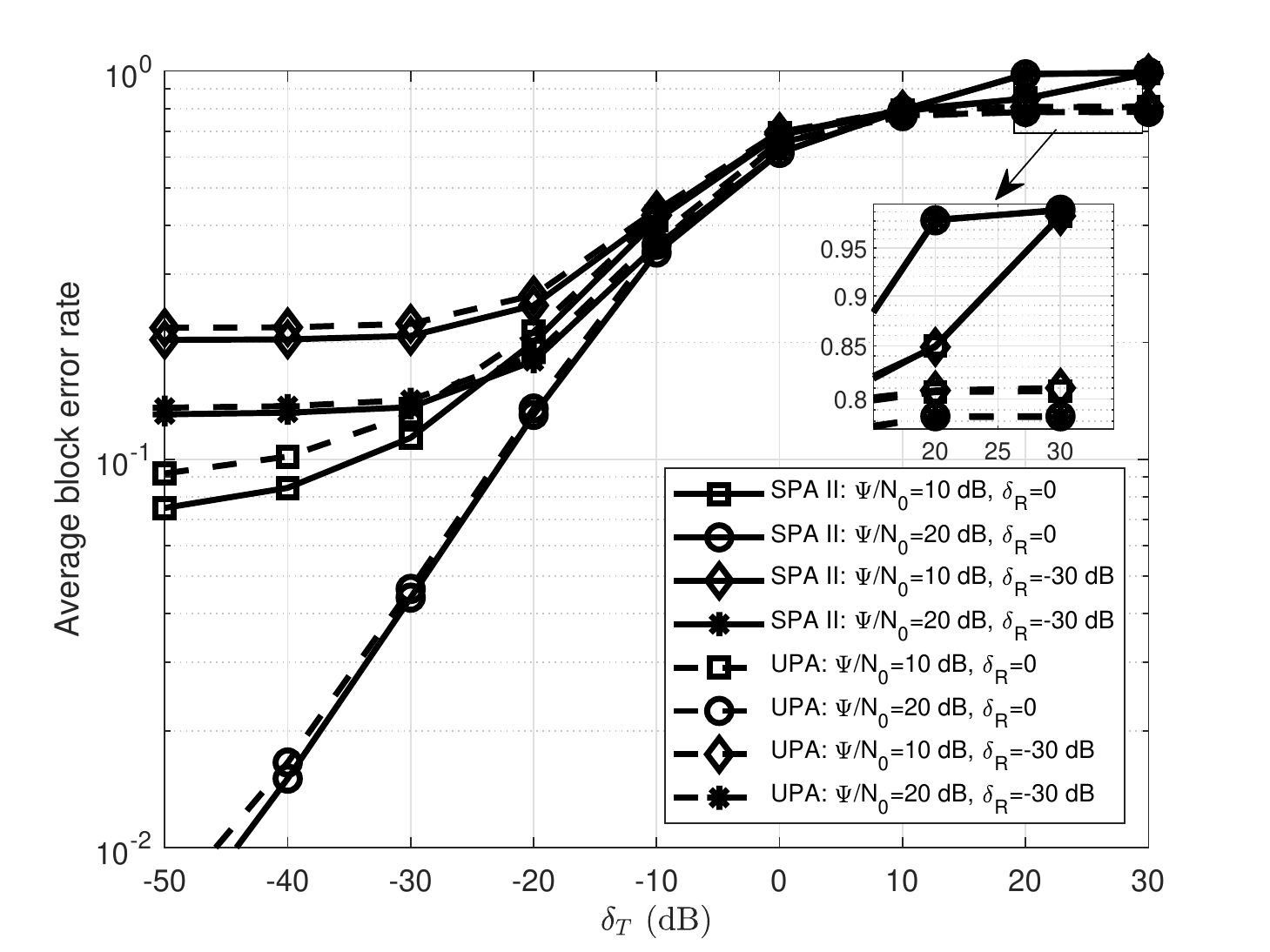}
        \caption{Imperfect CSI at the transmitter}
    \end{subfigure}%
~
    \begin{subfigure}[t]{0.5\textwidth}
        \centering
        \includegraphics[width=3.5in]{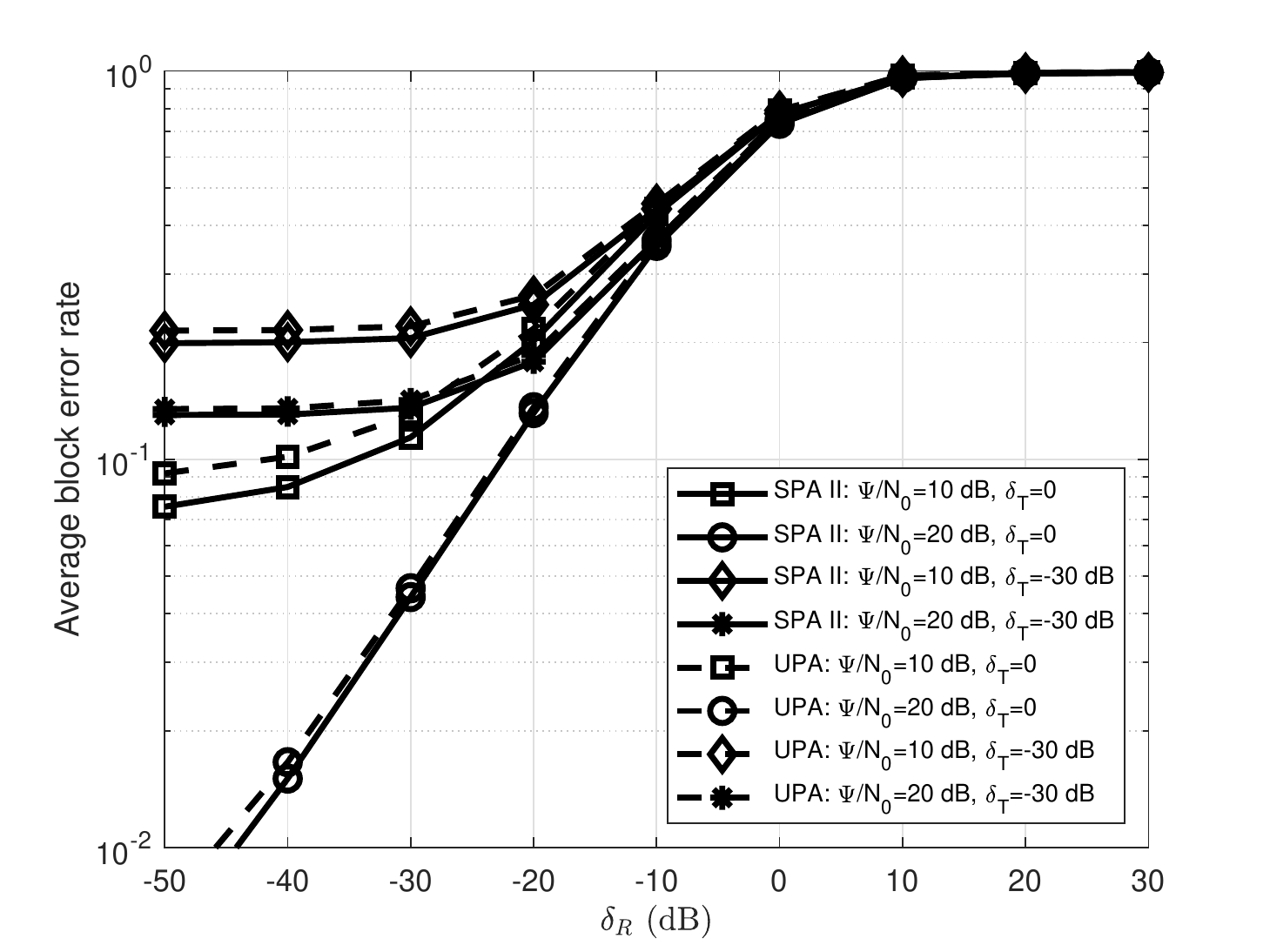}
        \caption{Imperfect CSI at receiver}
    \end{subfigure}
    \caption{Average BLER vs. the variance of channel estimation errors for imperfect CSI at the transmitter and receiver.}
    \label{csisimu}
\end{figure*}

\subsection{Verification of the Proposed Power Allocation Schemes}

To verify and compare the efficiency of the power allocation schemes introduced in Section \ref{pas}, we select EHC-OFDM-IM systems with tree architectures a-I, a-V, b-I, and b-II as examples. Numerical results generated by Monte Carlo simulations are demonstrated in Fig. \ref{PAsim}. From this figure, it is clear that with perfect CSI, the CSI-based power allocation schemes outperform the CSI-free uniform power allocation scheme when $\Psi$ is large. In addition, the two proposed suboptimal power allocation schemes provide almost the same outage performance, and the suboptimal power allocation scheme I is slightly worse than the suboptimal power allocation scheme II, because the optimization accuracy is lost when performing iterative algorithms. Hence, through these numerical results, the effectiveness and efficiency of the proposed power allocation schemes with a large power budget are substantiated.

\subsection{Impact of Imperfect CSI on Error Performance}
CSI is the key for implementing EHC-OFDM-IM systems, which is required at both transmitter and receiver for transmission adaptation and received OFDM block estimation. However, in practical situations, CSI might not be perfectly estimated, especially when parallel frequency channels are dynamic. In this subsection, we numerically evaluate the impact of imperfect CSI on error performance. We model the estimated CSI at the transmitter and receiver by the additive errors as follows: $\mathbf{H}_T=\mathbf{H}+\mathbf{E}_T$ and $\mathbf{H}_R=\mathbf{H}+\mathbf{E}_R$, where $\mathbf{E}_T$ and $\mathbf{E}_R$ are the diagonal matrices of channel estimation errors at the transmitter and receiver. The diagonal entries of $\mathbf{E}_T$ and $\mathbf{E}_R$ follow the complex Gaussian distribution $\mathcal{CN}(0,\delta_T)$ and $\mathcal{CN}(0,\delta_R)$, respectively  \cite{8734877}. 

Without loss of generality, we investigate the impact of imperfect CSI at the transmitter and receiver on EHC-OFDM-IM systems employing the tree architecture a-I, given $N=4$, $K=2$ and $M=2$ for the succinctness of illustration. Meanwhile, the suboptimal power allocation scheme II proposed in Section \ref{dsahdkj2hjdsreducpowsec} and the tailored uniform power allocation scheme introduced in Section \ref{opassstsc} are selected as examples for illustration. The simulation results are presented in Fig. \ref{csisimu}, from which we can discover several key impacts of CSI on the error performance of EHC-OFDM-IM systems. First, when the estimation error becomes severe, whether at the transmitter or the receiver, the coding gain brought by the CSI-based power allocation scheme diminishes rapidly, which is caused by the mismatch of codebooks for encoding and decoding at the transmitter and the receiver. As the codebook mismatch is reciprocal, we can see that there exists symmetry between Fig. \ref{csisimu}(a) and Fig. \ref{csisimu}(b) when $\delta_T$ and $\delta_R$ are small. 

When $\delta_T$ and $\delta_R$ become large, the CSI-free uniform power allocation scheme outperforms the CSI-based optimized power allocation schemes, since the erroneously estimated CSI only provides a misleading decision-making reference and yields a destructive impact on coded systems. Further, as used for detection purposes, the quality of CSI at the receiver dominates the error performance. Therefore, when $\delta_R$ becomes large, the average BLERs corresponding to all cases converge to a value close to unity. In other words, when $\delta_R$ is large, the estimation process at the receiver is equivalent to a random guess, which thereby can hardly provide a correct estimation of the received OFDM block. On the contrary, when $\delta_T$ is large and the uniform power allocation is adopted, there is still a possibility that the received OFDM block can be correctly decoded as long as the CSI at the receiver is accurate, even though a codebook mismatch exists.

\section{Conclusion}\label{c}
In this paper, we proposed the EHC-OFDM-IM scheme for multi-carrier systems, which is jointly enhanced by Huffman coding, lexicographic codebook design, and a dynamic power allocation strategy. The proposed EHC-OFDM-IM scheme is capable of utilizing all legitimate SAPs, adapting the bijective mapping relation between SAPs and leaves on a given Huffman tree and allocating transmit power to each active subcarrier by the CSI at the transmitter. In this way, the error performance is improved. To provide insights into the proposed system, we approximated the average BLER in closed form when the power allocation is independent of CSI. Then, we proposed two CSI-based dynamic power allocation schemes to minimize BLER. The first one utilizing the exact expression of the Q-function resulted in a transcendental equation set that requires high computational complexity for obtaining numerical solutions. The second scheme inspired by the exponential approximation of the Q-function is a reduced version of the first one, by which closed-form analytical solutions were provided. Numerical simulations verified our error performance analysis and the effectiveness of the two proposed power allocation schemes and revealed that the depth of the Huffman tree has a significant impact on the error performance when the SAP-to-leaf mapping relation is optimized based on CSI. In addition, because CSI is the key for both transmission adaptation and signal estimation, we numerically investigated the impacts of imperfect CSI at the transmitter and receiver on the error performance of EHC-OFDM-IM systems. Overall, this work provides an important step in OFDM-IM research.

\section*{Acknowledgment}
We thank the editor and the anonymous reviewers for the constructive comments, which have helped us improve the quality of the paper.

\bibliographystyle{IEEEtran}
\bibliography{bib}

\end{document}